\documentclass[aps,prx,twocolumn,superscriptaddress]{revtex4-2}
\usepackage[a4paper,margin=1.25cm]{geometry}
\usepackage{times}
\usepackage{graphicx}
\usepackage[font={small}]{caption}
\usepackage{caption}
\usepackage{subcaption}
\usepackage{amsmath}
\usepackage{amssymb}
\usepackage{tabularx}
\usepackage{placeins}
\usepackage{multirow}
\usepackage{braket}
\usepackage{color}
\usepackage{soul}
\usepackage{hyperref}
\usepackage[normalem]{ulem}
\usepackage{breakcites}
\usepackage{adjustbox}
\usepackage{mathtools}
\usepackage{float}
\usepackage{adjustbox}
\usepackage{dcolumn}% Align table columns on the decimal point
\usepackage{bm}% bold math
\usepackage{comment}
\usepackage{titlesec}
\usepackage{booktabs}
\usepackage{siunitx}
\usepackage{comment}
\usepackage{wrapfig}
\usepackage{changepage}
\renewcommand{\arraystretch}{1.1}
\usepackage[dvipsnames]{xcolor}

\titleformat{\subsubsection}{\normalfont\centering}{\thesubsubsection.}{1em}{}
\begin{document}
%~~~~~~~~~~~~~~~~~~~~~~~~~~~~~~~
\title{Accurate Atomic Correlation Energies and Virial Trends for Correlation-Consistent Effective Core Potentials III: Quantitative Separation of Fixed-Node Biases and Localization Errors in Cs–Rn} 
%~~~~~~~~~~~~~~~~~~~~~~~~~~~~~~~
\author{Omar Madany}
\author{Lubos Mitas}
\affiliation{Department of Physics, North Carolina State University, Raleigh, North Carolina 27695-8202, USA}
%\email{lmitas@ncsu.edu}
%~~~~~~~~~~~~~~~~~~~~~~~~~~~~~~~
\begin{abstract}
\noindent
\hspace{-1.5cm}%
\begin{minipage}{\dimexpr\textwidth-2.0cm\relax}
\begin{wrapfigure}{r}{0.55\textwidth}
    \vspace{0cm}
    \hspace{-0.5cm}
    \centering
    \includegraphics[
        width=0.57\textwidth
    ]{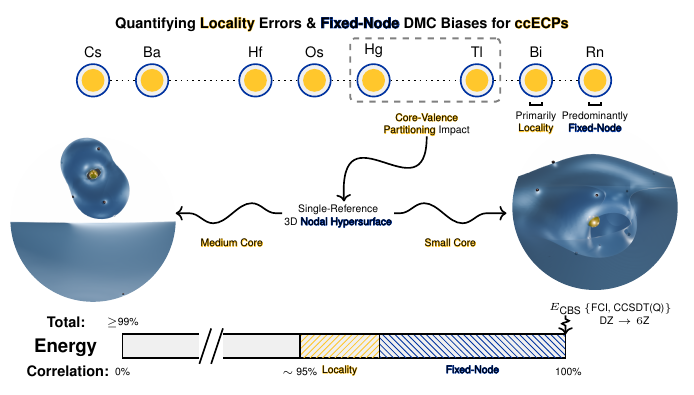}
    \vspace{-0.5cm}
\end{wrapfigure}
Atomic benchmarks for sixth-period elements (Cs--Rn) are established to sub-millihartree precision under correlation-consistent effective core potentials (ccECPs) using coupled-cluster [up to CCSDT(Q)] and full configuration interaction methods. These references systematically characterize electron correlation, potential-to-kinetic energy partitioning, and combined fixed-node and locality biases ($\eta$) in fixed-node diffusion Monte Carlo (FNDMC). The valence kinetic-to-total energy ratio ($\xi$) departs from the non-relativistic all-electron Coulombic virial limit ($\xi=100\%$), remaining confined between 25.8\% and 37.6\%. The Un-pseudizing of semi-core shells decreases $\xi$ as the attractive potential energy deepens faster than the accompanying kinetic energy escalates. Periodic trends in $\xi$ reveal a $5d$ plateau near 34\%, with a lanthanide-contraction-driven maximum in Hf, followed by a sharp decline across the $6p$ block. In FNDMC, $\eta$ escalates rapidly at the onset of group-1/2 and toward the $6p$ closed-shell limit, peaking at 5.42\% in Rn. Although fixed-node errors dominate $\eta$ in most sixth-period elements, a locality error inversion occurs in Bi, with Pb exhibiting comparable error magnitudes. Furthermore, core-valence partitioning alters trial wavefunction nodal topology: in Tl, three-dimensional nodal cross-sections reveal isolated toroidal domains that confine walkers. This work completes the periodic-table-wide ccECP database from H to Rn, revealing how pseudopotential choices govern energy partitioning, nodal topology, and the balance between fixed-node and localization errors.
\end{minipage}
\end{abstract}
%%%~~~~~~~~~~~~~~~~~~~~~~~~~~~~~~~~~~~~~~~~~~~~~~~~~~~
\maketitle
%%%~~~~~~~~~~~~~~~~~~~~~~~~~~~~~~~~~~~~~~~~~~~~~~~~~~~
\section{Introduction}
\label{intro}
Sixth-period elements ($Z=55\text{--}86$) require careful relativistic treatment, as massive nuclear charges drive core electrons to relativistic velocities and render nonrelativistic quantum mechanics insufficient~\cite{Schwerdtfeger2011}. Relativistic effects alter the spatial topology of the wavefunction, driving orbital contraction and stabilization of the valence shell~\cite{Pyykko2002, Dolg2011}. This effect, compounded by the lanthanide contraction across the filled $4f$ shell, localizes and concentrates electron density~\cite{Paul1975,Kuchle1997}. This high local density influences wavefunction curvature and introduces nodal non-linearities, shaping the topological complexity of the many-body nodal surfaces in real-space quantum Monte Carlo (QMC) methods~\cite{kevin2014}.

Accurate modeling of electron correlation in these heavy systems is essential but computationally demanding~\cite{szabo2012}. In real-space QMC, all-electron (AE) treatments become computationally intractable due to severe core-energy fluctuations that drive computational time scaling as $\mathcal{O}(Z^{5.5-6.5})$~\cite{Ceperley1986, Hammond1987}. To circumvent this bottleneck, valence-only Hamiltonians are utilized by introducing effective core potentials (ECPs) or pseudopotentials. The effective core charge $Z_{\text{eff}}$ specifies the core-valence boundary, defining the extent of the explicit core-valence correlation. While traditional pseudopotentials were developed within mean-field settings, correlation-consistent ECPs (ccECPs) are constructed from a many-body framework, explicitly parameterized to reproduce many-body core-valence and core-core correlation effects~\cite{Bennett2017,Bennett2018,Annaberdiyev2018,Wang2019,Wang2022,Kincaid2022,annaberdiyev2023,Haihan2024,Omar2025,Omar2026,pseudopotentiallibrary}. Nevertheless, because all semi-local ECPs share the same non-local angular momentum projection structure, the physical mechanisms governing wavefunction curvature and kinetic-energy partitioning are expected to be qualitatively universal across ECP choices~\cite{Schwerdtfeger2011,Dzubak2017,acc_engI}.

Even within a valence-only Hamiltonian, projecting the ground state in fixed-node diffusion Monte Carlo (FNDMC) is complicated by the non-local sign problem, which arises from the angular momentum projection operators in the pseudopotential~\cite{foulkes2001,Krogel2017,Dzubak2017}. Standard FNDMC calculations typically resolve this using the non-variational locality approximation (LA)~\cite{Hurley1987,Mitas1991}, which is computationally efficient and yields a localization (locality) error scaling as the square of the trial wavefunction error. Alternatively, the variational T-moves (TM) scheme restores the energy upper bound and ensures algorithmic stability by transferring sign-violating off-diagonal projection elements to the diagonal~\cite{casula2006, casula2010, Haihan2022}. Regardless of the scheme chosen, the resulting energies remain highly sensitive to the quality of the trial wavefunction, making the systematic characterization of these localization biases a critical step for high-precision real-space QMC.

Effective core potentials also alter potential-to-kinetic energy partitioning. While non-relativistic all-electron Hamiltonians satisfy the Coulombic virial theorem, pseudopotential Hamiltonians obey a modified relation accounting for the kinetic energy reduction near atomic nuclei caused by core elimination~\cite{Ihm1979,Pickett1989,Bitter2010,Schwerdtfeger2011}. Consequently, quantifying this deviation through the kinetic-to-total energy ratio, $\xi = E^{\text{kin}}/|E^{\text{total}}|$, provides a rigorous metric to assess how pseudopotential core boundaries drive the global energy partitioning and wavefunction curvature.

The magnitude and behavior of the fixed-node bias are connected to local electron density and node non-linearity. Although studies on lighter elements show that high local density and nodal distortions amplify fixed-node errors~\cite{Kevin2012,kevin2014}, sixth-period elements present a greater challenge because relativistic spatial contraction concentrates valence density near the core boundary. These nodal geometries depend significantly on the core-valence partitioning of the pseudopotential. Furthermore, because semi-local ECPs require localization approximations that depend on the trial wavefunction, it remains quantitatively unclear whether heavy-element DMC biases are dominated by these localization errors or by fixed-node constraints. Resolving this requires reference energies from correlated many-body approaches at the complete basis set (CBS) limit to eliminate basis-set incompleteness errors~\cite{Varandas2007,szabo2012}. Extrapolating coupled-cluster energies [up to CCSDT(Q)] and full configuration interaction (FCI) establishes an exact or near-exact valence baseline, enabling the separation of fixed-node ($\epsilon_{\text{FN}}$) and localization ($\epsilon_{\text{loc}}$) biases across the Cs--Rn series.
 
Prior electronic structure benchmarks have developed along distinct lines, such as AE relativistic non-correlated baselines~\cite{Desclaux1973,Kotochigova1997,Rolfes2020}, AE relativistic correlated wave-function calculations utilizing relativistic coupled-cluster and perturbation theories~\cite{Ishikawa1994,Kumar2021,Hoyer2023,Dzuba2024}, and non-relativistic many-body methods including QMC and configuration-interaction limits~\cite{Kotochigova1997,Chakravorty1993,Luchow1996,Jitrik1997,Booth2010,Cleland2011,McCarthy2011,McCarthy2012,Buendia2013,Ranasinghe2013,Scemama2014,Ranasinghe2015,Li2019,Peng2023}. Despite these advances, a systematic, self-consistent database pairing accurate many-body CBS limits with real-space DMC nodal analyses and energy partitioning remains entirely absent for the sixth-period elements. In this work, we address this gap by establishing highly accurate total, correlation, and kinetic energy benchmarks for the Cs--Rn atoms. Using high-level coupled-cluster and configuration-interaction methods with systematic CBS extrapolations alongside QMC, we dissect potential-to-kinetic energy partitioning ($\xi$), explore the topological complexity of single-reference nodal surfaces, and partition fixed-node and localization errors. 

As Hartree–Fock orbitals form the standard trial wavefunction ansatz in routine QMC applications, establishing the exact fixed-node bias of these single-determinant Slater–Jastrow nodal surfaces provides the reference baseline required before applying costly multi-determinant or orbital optimizations. This study completes our systematic, periodic-table-wide benchmarking efforts succeeding our prior studies on H--Kr~\cite{acc_engI} and Rb--Xe~\cite{acc_engII}, thereby delivering a unified database of many-body ccECP references across the first six periods of the periodic table.
\section{Methods}
\label{sec:methods}
%%%~~~~~~~~~~~~~~~~~~~~~~~~~~~~~~~~~~~~~~~~~~~~~~~~~~~
\subsection{Non-Local Sign Problem and Localization Schemes}
\label{subsec:ccECP}

The ccECP averaged relativistic effective potential (AREP) operator $\hat{V}^{\text{AREP}}$ is introduced to replace the core electrons, reducing the many-body problem to the valence space of $Z_{\text{eff}}$ electrons. This valence AREP is expressed as a sum of single-particle operators over all valence electrons:

\begin{equation}
\label{Varepform}
\hat{V}_{\text{AREP}} = \sum_{i=1}^{Z_{\text{eff}}} \left[ V_{\text{loc}}(r_i) + \sum_{\ell=0}^{\ell_{\text{max}}} \Delta V_\ell^{\text{AREP}}(r_i) \hat{P}_\ell^{(i)} \right]
\end{equation}

where $V_{\text{loc}}(r_i)$ is the local potential, $\Delta V_\ell^{\text{AREP}}(r_i)$ represents the radial component of the non-local potential that accounts for the $\ell$-dependent interaction of the valence electrons with the core, and $\hat{P}_\ell^{(i)}$ is the projection operator which acts purely on the angular coordinates of electron $i$. 

In real-space DMC, non-local angular-momentum projection operators in semi-local Hamiltonians generate off-diagonal transitions in the imaginary-time propagator. We employ the T-moves localization scheme~\cite{casula2006,casula2010,kim2018,paul2020} which partially localizes the pseudopotential by forcing sign-violating off-diagonal terms onto the diagonal while executing valid terms as spatial transitions. Compared to the standard localization approximation, the T-moves scheme also significantly reduces the sensitivity of the projected energy to the Jastrow parametrization~\cite{Haihan2022,jaehyung2015}.

%%%~~~~~~~~~~~~~~~~~~~~~~~~~~~~~~~~~~~~~~~~~~~~~~~~~~~
\subsection{Reference \textit{Ab Initio} Energies and Complete Basis Set Extrapolations}
\label{subsec:ref_cbs}

To establish exact or near-exact reference benchmarks under the valence-only ccECP Hamiltonians, valence correlation energies were evaluated using a coupled cluster hierarchy [CCSD(T), CCSDT, and CCSDT(Q)], CISD, and FCI in \texttt{MOLPRO} and \texttt{MRCC} packages~\cite{molpro-2012,molpro-2020,molpro-3,molpro-cc,molpro-seward,molpro-mcscf1,molpro-mcscf2,molpro-fci,mrcc-1,mrcc-2}. To systematically minimize basis set truncation errors, Hartree-Fock and correlation energies were extrapolated separately to the CBS limit using sequences of correlation-consistent basis sets, aug-cc-p(C)V$n$Z~\cite{acc_engI,acc_engII,Varandas2007}. For highly correlated methods that were computationally intractable at the largest basis sets, the high-order correlation energy contribution was estimated using a ratio-based scaling procedure described in Refs.~\cite{acc_engI,acc_engII}. The resulting extrapolated and estimated contributions were then used to obtain the corresponding CBS-limit energies.

The expectation value of the kinetic energy was evaluated from correlated many-body wavefunctions. Since FCI calculations are restricted by the factorial scaling of the active space, CISD was used as a baseline where necessary. The missing FCI kinetic energies at larger basis sets were estimated using an analogous FCI/CISD scaling relation to that established in Refs.~\cite{acc_engI,acc_engII}. The kinetic energy was subsequently extrapolated to the CBS limit. Differences between results obtained with the largest basis sets are utilized to estimate extrapolation uncertainties, with detailed extrapolation formulas provided in the Supplementary Information.

To characterize the spatial confinement and quantify the influence of the ECP on the potential-to-kinetic energy partitioning, we evaluate the normalized kinetic-energy ratio:
\begin{equation}
\label{eq:xi}
\xi~(\%)= \frac{E_{\text{CBS}}^{\text{kin}}}{|E_{\text{CBS}}^{\text{total}}|}\times100\%
\end{equation}
Both $E_{\text{CBS}}^{\text{kin}}$ and $E_{\text{CBS}}^{\text{total}}$ represent valence-only expectation values evaluated under the effective valence Hamiltonian. For an exact, non-relativistic AE Coulomb system, the standard virial theorem requires $2\langle \hat{T} \rangle = -\langle \hat{V} \rangle$, which yields $\xi = 100\%$. Assessing $\xi$ reveals how core-valence separation and non-local projection operators alter kinetic and potential energy partitioning in the core region.

%%%~~~~~~~~~~~~~~~~~~~~~~~~~~~~~~~~~~~~~~~~~~~~~~~~~~~
\subsection{Quantum Monte Carlo and the Associated Biases}
\label{subsec:qmc}

The real-space FNDMC method projects the many-body ground-state component from a trial wavefunction by propagating configurations in imaginary time using a Green's function propagator~\cite{foulkes2001}. To eliminate finite time-step bias, we evaluate DMC energies across a sequence of imaginary time-steps ($\tau \in \{0.02, 0.01, 0.005, 0.0025\}\ \text{Ha}^{-1}$) and extrapolate to the zero-timestep limit ($\tau \to 0$) using a weighted least-squares linear regression.

Our single-reference trial wavefunctions are constructed using Hartree-Fock orbitals obtained from \texttt{PySCF}~\cite{pyscf1,pyscf2,pyscf3,pyscf4} or \texttt{GAMESS}~\cite{GAMESS} augmented by optimized Jastrow terms~\cite{kim2018,paul2020}: 
\begin{equation}
\Psi_T(\mathbf{R}) = e^{J(\mathbf{R})} D^{\uparrow}(\mathbf{R}_{\uparrow}) D^{\downarrow}(\mathbf{R}_{\downarrow})
\end{equation}
where $D^{\uparrow}$ and $D^{\downarrow}$ are Slater determinants of single-particle Hartree-Fock orbitals, and $J(\mathbf{R})$ is the Jastrow correlation factor. 
Because the Jastrow factor is strictly positive-definite, the nodal hypersurface is dictated entirely by the single-determinant product. Consequently, the extracted fixed-node errors ($\epsilon_{\text{FN}}$) measure the topological limitation of the single-configuration Hartree–Fock nodal boundary. To capture dynamic correlation and reduce local energy variance, the Jastrow factor is partitioned into one-, two-, and three-body terms ($J_1$, $J_2$, and $J_3$, corresponding to electron-nucleus, electron-electron, and electron-electron-nucleus interactions, respectively). The Jastrow factor parameters are variationally optimized using the energy and variance minimization protocols in \texttt{QMCPACK}~\cite{kim2018,paul2020}. Walker populations are kept sufficiently large ($\ge 2000$ walkers) to render population control biases negligible.

To evaluate overall FNDMC accuracy against our CBS reference benchmarks, we define the total absolute energy error ($\epsilon$) and the normalized relative correlation error ($\eta$):
\begin{equation}
    \epsilon= |E_{\text{CBS}}^{\text{total}}-E_{\text{$\tau \to$ 0}}^{\text{DMC,total}}|
\end{equation}

\begin{equation}\label{eq:eta}
    \eta~(\%)= \frac{\epsilon}{|E_{\text{CBS}}^{\text{corr}}|}\times100\%
\end{equation}

To determine whether fixed-node single-configuration Hartree-Fock nodal topology or pseudopotential localization controls this overall bias, we partition $\epsilon$ into fixed-node ($\epsilon_{\text{FN}}$) and locality ($\epsilon_{\text{loc}}$) components. Following the formalism of Refs.~\cite{casula2005,Krogel2017,Dzubak2017}, systematically modifying the Jastrow factor alters only the localization error while preserving a constant fixed-node error. By evaluating the FNDMC energies across sequential Jastrow levels (no-Jastrow, J12, and J123), we parameterize a linear regression model of the DMC energy as a function of the corresponding VMC energy. Extrapolation to the limit $E_{\text{VMC}} - E_{\text{DMC}} \to 0$ yields the locality-error-corrected limit ($E_{\text{LEC}}$), which corresponds to the exact eigenvalue of the nonlocal ECP Hamiltonian restricted to the nodal surface of $\Psi_T$.

Using this limit, we define the absolute locality error ($\epsilon_{\text{loc}}$) as the difference between the extrapolated zero-timestep J123-DMC energy and the locality-corrected limit:
\begin{equation}
\label{eps_loc}
\epsilon_{\text{loc}} = |E_{\tau \to 0}^{\text{DMC,total}} - E_{\text{LEC}}|
\end{equation}

which isolates the energetic penalty incurred by the localization approximation (T-moves). The remaining absolute fixed-node error ($\epsilon_{\text{FN}}$) is subsequently defined as the difference between the exact CBS reference and the locality-corrected energy:
\begin{equation}
\label{eps_fn}
\epsilon_{\text{FN}} = |E_{\text{CBS}}^{\text{total}} - E_{\text{LEC}}|
\end{equation}
which isolates the topological error inherent to the fixed-node approximation. To assess their relative impacts on correlation recovery, we normalize these quantities against the CBS valence correlation energy:

\begin{equation}
\label{eq:eta_loc}
\eta_{\mathrm{loc}}~(\%) = \frac{\epsilon_{\mathrm{loc}}}{|E_{\mathrm{CBS}}^{\mathrm{corr}}|} \times 100\%
\end{equation}

\begin{equation}
\label{eq:eta_FN}
\eta_{\mathrm{FN}}~(\%) = \frac{\epsilon_{\mathrm{FN}}}{|E_{\mathrm{CBS}}^{\mathrm{corr}}|} \times 100\%
\end{equation}
This separation isolates the relative impact of each error component across the sixth-period series.

%%%~~~~~~~~~~~~~~~~~~~~~~~~~~~~~~~~~~~~~~~~~~~~~~~~~~~
\section{Results and Discussion}
\label{sec:results}

%%%===================================================%%%
%%%~~~~~~~~~~~~~~~~~~~~~~~~~~~~~~~~~~~~~~~~~~~~~~~~~~~%%%
%%%===================================================%%%

\subsection{Exact and Near-Exact Correlated Reference Energies}
\label{subsec:exact_ref_energy}
Tables \ref{tab:compact_Total_Energy_corr_methods_SA_1}--\ref{tab:compact_Total_Energy_corr_methods_SA_2} present extrapolated CBS total and valence correlation energies, with extrapolation uncertainties consistently below $1$~mHa. For Tl ($Z_{\text{eff}}=3$) and Pb ($Z_{\text{eff}}=4$), CCSDT(Q) recovers $>99.9\%$ of the valence correlation energy, differing from FCI CBS limits by $0.14$ mHa and $0.06$ mHa, respectively. CCSDT(Q) therefore serves as the exact valence benchmark surrogate for high-$Z_{\text{eff}}$ elements where FCI is computationally intractable~\cite{szabo2012}.

\begin{table}[htbp!]
\setlength{\tabcolsep}{1pt}
\centering
\small
\caption{Extrapolated CBS valence total energies and correlation energies for the sixth-period elements (Cs--Au). Calculations are performed under ccECP Hamiltonians utilizing a Coupled Cluster hierarchy [CCSD(T), CCSDT, and CCSDT(Q)] and FCI as described in Sec. \ref{subsec:ref_cbs} with statistical uncertainties from the extrapolation fits are enclosed in parentheses. Additional tables containing all finite-basis $E_{nZ}$ energies are provided in the Supplementary Information.}
\label{tab:compact_Total_Energy_corr_methods_SA_1}
\begin{tabular}{lcccccc}
\hline
\textbf{Method} & \textbf{Atom} & \textbf{$\mathbf{Z_{\text{eff}}}$} & $E_{\rm CBS}$ (Ha) & \textbf{Atom} & \textbf{$\mathbf{Z_{\text{eff}}}$} & $E_{\rm CBS}$ (Ha) \\
\hline
CISD & \multirow{6}{*}{\textbf{Cs}} & \multirow{6}{*}{\textbf{9}} & -0.228777(61) & \multirow{6}{*}{\textbf{Ba}} & \multirow{6}{*}{\textbf{10}} & -0.27640(57) \\
CCSD(T) & & & -0.25157(17) & & & -0.31589(59) \\
CCSDT & & & -0.25187(14) & & & -0.31599(66) \\
CCSDT(Q) & & & -0.25230(16) & & & -0.31689(62) \\
HF & & & -19.8645954(41) & & & -25.172548(49) \\
Total & & & -20.11690(16) & & & -25.48944(62) \\
\\
CISD & \multirow{6}{*}{\textbf{Hf}} & \multirow{6}{*}{\textbf{12}} & -0.34896(83) & \multirow{6}{*}{\textbf{Ta}} & \multirow{6}{*}{\textbf{13}} & -0.38588(86) \\
CCSD(T) & & & -0.39211(90) & & & -0.43511(98) \\
CCSDT & & & -0.39251(87) & & & -0.43578(95) \\
CCSDT(Q) & & & -0.39355(87) & & & -0.43710(95) \\
HF & & & -48.01040(35) & & & -56.856219(13) \\
Total & & & -48.40396(94) & & & -57.29332(95) \\
\\
CISD & \multirow{6}{*}{\textbf{W}} & \multirow{6}{*}{\textbf{14}} & -0.42964(62) & \multirow{6}{*}{\textbf{Re}} & \multirow{6}{*}{\textbf{15}} & -0.47857(67) \\
CCSD(T) & & & -0.48753(67) & & & -0.54102(73) \\
CCSDT & & & -0.48803(66) & & & -0.54014(70) \\
CCSDT(Q) & & & -0.48973(66) & & & -0.54136(76) \\
HF & & & -66.6938184(43) & & & -77.774279(33) \\
Total & & & -67.18355(66) & & & -78.31564(76) \\
\\
CISD & \multirow{6}{*}{\textbf{Os}} & \multirow{6}{*}{\textbf{16}} & -0.5454(13) & \multirow{6}{*}{\textbf{Ir}} & \multirow{6}{*}{\textbf{17}} & -0.60583(75) \\
CCSD(T) & & & -0.6212(14) & & & -0.69119(69) \\
CCSDT & & & -0.6200(14) & & & -0.6872(69) \\
CCSDT(Q) & & & -0.6214(14) & & & -0.6822(28) \\
HF & & & -90.019562(17) & & & -103.6650439(12) \\
Total & & & -90.6410(14) & & & -104.3472(28) \\
\\
CISD & \multirow{6}{*}{\textbf{Pt}} & \multirow{6}{*}{\textbf{18}} & -0.6671(15) & \multirow{6}{*}{\textbf{Au}} & \multirow{6}{*}{\textbf{19}} & -0.7207(14) \\
CCSD(T) & & & -0.7557(15) & & & -0.8163(14) \\
CCSDT & & & -0.7538(15) & & & -0.8140(14) \\
CCSDT(Q) & & & -0.7550(15) & & & -0.8150(14) \\
HF & & & -118.754596(87) & & & -135.0061224(26) \\
Total & & & -119.5096(15) & & & -135.8212(14) \\
\end{tabular}
\end{table}

\begin{table}[htbp!]
\setlength{\tabcolsep}{1pt}
\centering
\small
\caption{Extrapolated CBS valence total and correlation energies for the sixth-period elements (Hg--Rn). Methodologies, CBS definitions, and formatting conventions are identical to those described in Table \ref{tab:compact_Total_Energy_corr_methods_SA_1}.}
\label{tab:compact_Total_Energy_corr_methods_SA_2}
\begin{tabular}{lcccccc}
\hline
\textbf{Method} & \textbf{Atom} & \textbf{$\mathbf{Z_{\text{eff}}}$} & $E_{\rm CBS}$ (Ha) & \textbf{Atom} & \textbf{$\mathbf{Z_{\text{eff}}}$} & $E_{\rm CBS}$ (Ha) \\
\hline
CISD & \multirow{6}{*}{\textbf{Hg}} & \multirow{6}{*}{\textbf{12}} & -0.6093(25) & \multirow{6}{*}{\textbf{Hg}} & \multirow{6}{*}{\textbf{20}} & -0.78767(37) \\
CCSD(T) & & & -0.7006(17) & & & -0.90117(14) \\
CCSDT & & & -0.6960(18) & & & -0.898208(89) \\
CCSDT(Q) & & & -0.6997(19) & & & -0.899644(87) \\
HF & & & -41.8227544(73) & & & -152.53125143(75) \\
Total & & & -42.5224(19) & & & -153.430896(87) \\
\\
CISD & \multirow{6}{*}{\textbf{Tl}} & \multirow{6}{*}{\textbf{13}} & -0.6235(17) & \multirow{6}{*}{\textbf{Tl}} & \multirow{6}{*}{\textbf{21}} & -0.8160(24) \\
CCSD(T) & & & -0.7117(10) & & & -0.9273(33) \\
CCSDT & & & -0.7078(12) & & & -0.9247(32) \\
CCSDT(Q) & & & -0.7110(12) & & & -0.9259(32) \\
HF & & & -49.43913148(66) & & & -171.482291(80) \\
Total & & & -50.1502(12) & & & -172.4082(32) \\
\\
CISD & \multirow{7}{*}{\textbf{Tl}} & \multirow{7}{*}{\textbf{3}} & -0.04288(18) & \multirow{7}{*}{\textbf{Pb}} & \multirow{7}{*}{\textbf{4}} & -0.057667(16) \\
CCSD(T) & & & -0.04387(18) & & & -0.059847(17) \\
CCSDT & & & -0.04302(18) & & & -0.060037(19) \\
CCSDT(Q) & & & -0.04386(18) & & & -0.060064(19) \\
FCI & & & -0.04400(18) & & & -0.06012(10) \\
HF & & & -1.987898(66) & & & -3.40225828(16) \\
Total & & & -2.03190(19) & & & -3.46238(10) \\
\\
CISD & \multirow{7}{*}{\textbf{Bi}} & \multirow{7}{*}{\textbf{5}} & -0.073684(32) & \multirow{7}{*}{\textbf{Po}} & \multirow{7}{*}{\textbf{6}} & -0.11625(27) \\
CCSD(T) & & & -0.07757(10) & & & -0.12427(30) \\
CCSDT & & & -0.077673(22) & & & -0.12452(31) \\
CCSDT(Q) & & & -0.077737(21) & & & -0.12465(31) \\
FCI & & & -0.077745(21) & & & -0.12466(31) \\
HF & & & -5.327912(13) & & & -7.76217666(32) \\
Total & & & -5.405658(24) & & & -7.88684(31) \\
\\
CISD & \multirow{7}{*}{\textbf{At}} & \multirow{7}{*}{\textbf{7}} & -0.15712(31) & \multirow{7}{*}{\textbf{Rn}} & \multirow{7}{*}{\textbf{8}} & -0.19439(22) \\
CCSD(T) & & & -0.17016(34) & & & -0.21237(30) \\
CCSDT & & & -0.17039(34) & & & -0.21253(33) \\
CCSDT(Q) & & & -0.17062(34) & & & -0.21289(34) \\
FCI & & & -0.17064(34) & & & -0.21290(34) \\
HF & & & -10.7555427(47) & & & -14.6437544(14) \\
Total & & & -10.92618(34) & & & -14.85665(34) \\
\end{tabular}
\end{table}

In Os--Hg ($5d$ series), perturbative triples in CCSD(T) systematically overestimate valence correlation energy. This over-correlation is corrected by the fully iterative, non-perturbative triples treatment in CCSDT, while CCSDT(Q) recovers higher-order quadruples necessary to converge systematically to the exact limit. The physical manifestation of this core-polarization and triples overestimation is well-documented in the mercury dimer, where perturbative triples deepen the weak dispersion well, necessitating high-order iterative coupled-cluster treatments to achieve quantitative agreement with spectroscopic benchmarks~\cite{Pahl2010, petrov2010}. Using CCSD(T) references would distort the extracted FNDMC error partition ($\epsilon_{\text{FN}}$ and $\epsilon_{\text{loc}}$), making iterative CCSDT(Q) baselines necessary.

Table~\ref{tab:compact_Kinetic_Energy_corr_methods_SA} presents the extrapolated CBS kinetic energies ($E_{\rm CBS}^{\rm kin}$). For Tl ($Z_{\text{eff}}=3$) and Pb ($Z_{\text{eff}}=4$), CISD reaches the FCI valence kinetic energy limits to within $2.3$~mHa and $4.0$~mHa, respectively. Furthermore, $E_{\rm CBS}^{\rm kin}$ scales steeply with the correlated valence space size; expanding $Z_{\text{eff}}$ in thallium from $3$ to $13$ and $21$ increases $E_{\rm CBS}^{\rm kin}$ from $0.761$~Ha to $31.048$~Ha and $56.467$~Ha, respectively. This increase reflects the localized kinetic energy density of the $5s5p$ and $5d$ semi-core shells. Previous studies have attributed the magnitude of fixed-node errors in QMC evaluations to changes in nodal hypersurface complexity driven by this localized density~\cite{Kevin2012,kevin2014,mitas2025}.

\begin{figure}[!htbp]
\centering
\begin{subfigure}{0.49\textwidth}
    \includegraphics[width=\textwidth]{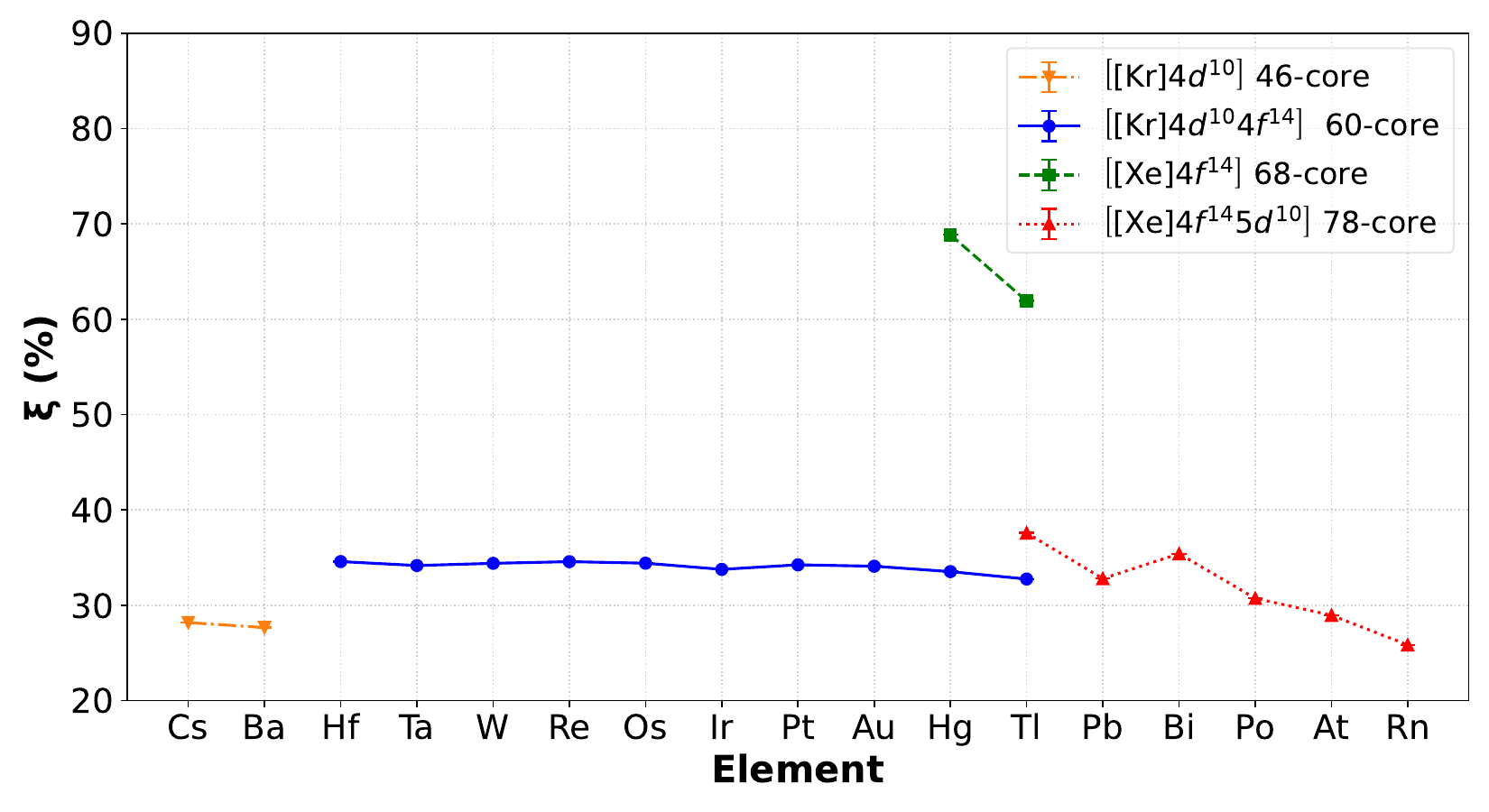}
    \caption{Ratio $\xi$ for the targeted Cs--Rn ccECPs as a function of the effective valence charge ($Z_{\text{eff}}$).}
    \label{fig:xi_6th_row}
\end{subfigure}\hfill
\begin{subfigure}{0.49\textwidth}
    \includegraphics[width=\textwidth]{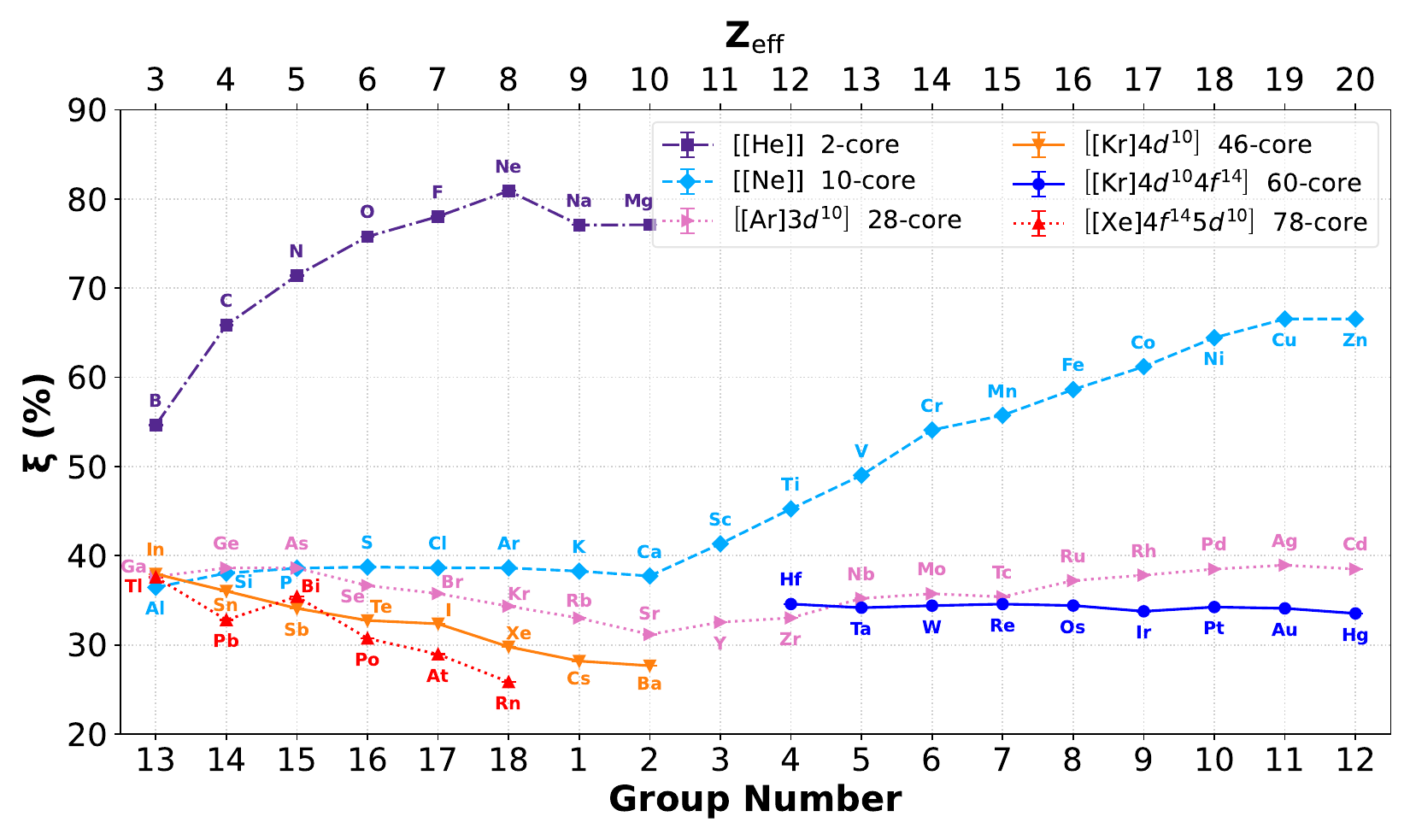}
    \caption{Extended periodic-table-wide comparison of the valence kinetic energy partitioning ratio $\xi$ across ccECP families.}
    \label{fig:xi_all_current_ccecps}
\end{subfigure}
\caption{Valence potential-to-kinetic energy partitioning ratio ($\xi$) evaluated as the ratio of the CBS kinetic energy to the magnitude of the CBS total energy. \textbf{(a)} Ratios ($\xi$) for the sixth-period (Cs--Rn) ccECPs as a function of the effective valence charge ($Z_{\text{eff}}$). \textbf{(b)} Comparison of $\xi$ across the periodic table, integrating the current Cs--Rn results with previously published H--Kr and Rb--Xe ccECPs [Refs.~\cite{acc_engI,acc_engII}], ordered by effective core charge ($Z_{\text{eff}}$), periodic group, and core size. Because core shells are projected out, both the kinetic and total energies represent valence-only properties of the semi-local Hamiltonian, driving the ratio $\xi$ far below the AE Coulombic limit. The calculation details are described in Sec.~\ref{subsec:exact_ref_energy}, with numerical values tabulated in Table~\ref{tab:compact_DMC_bias_Ratio_kinetic_corr_methods_SA}.}
\label{fig:xi_core_size_plots}
\end{figure}

\begin{figure*}[!htbp]
\centering
\begin{subfigure}{0.33\textwidth}
    \includegraphics[width=\textwidth]{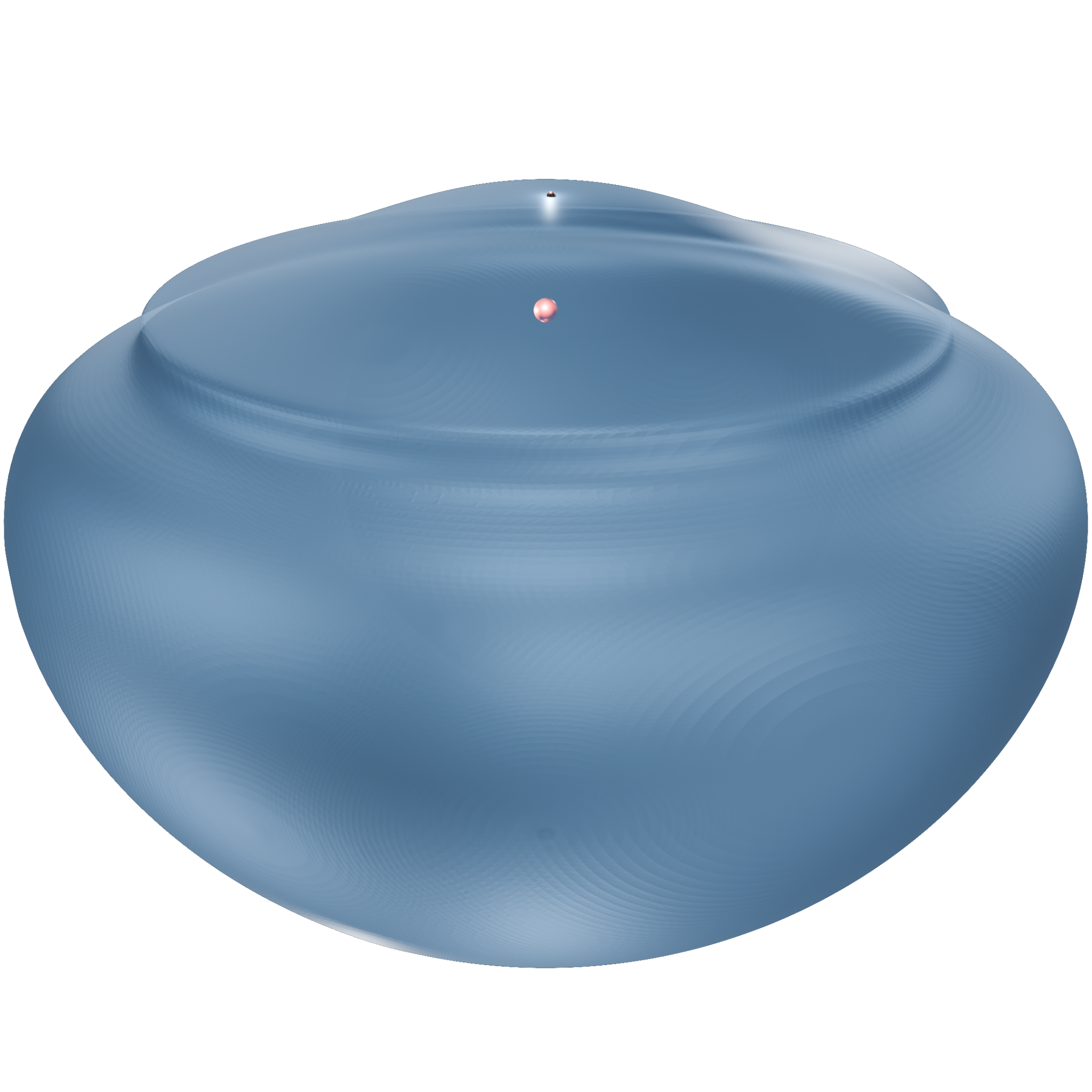}
    \caption{Nodal surface for the large-core Tl ccECP.}
    \label{fig:large_Tl_node}
\end{subfigure}\hfill
\begin{subfigure}{0.33\textwidth}
    \includegraphics[width=\textwidth]{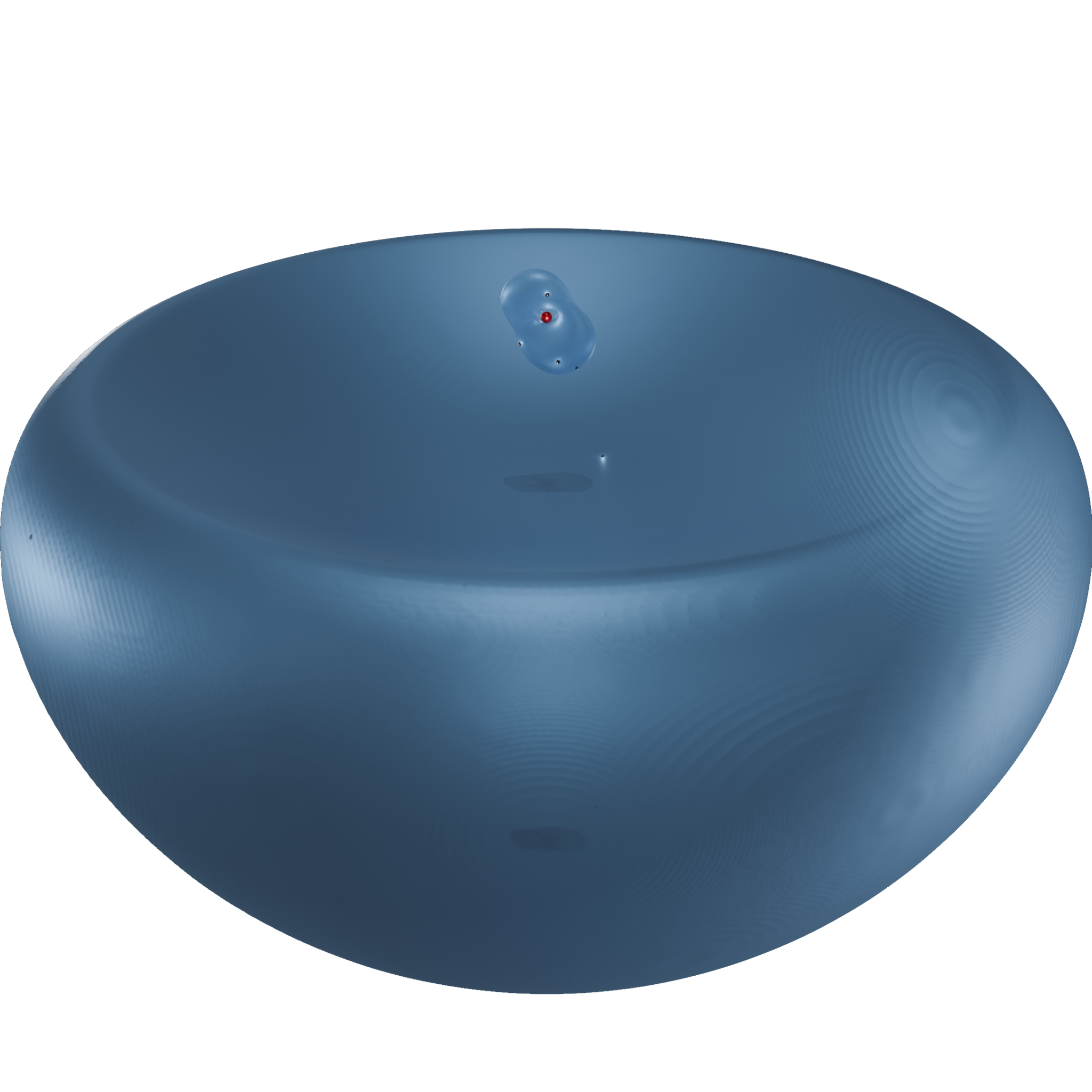}
    \caption{Nodal surface for the medium-core Tl ccECP.}
    \label{fig:medium_Tl_node}
\end{subfigure}\hfill
% \vspace{\baselineskip} 
\begin{subfigure}{0.33\textwidth}
    \includegraphics[width=\textwidth]{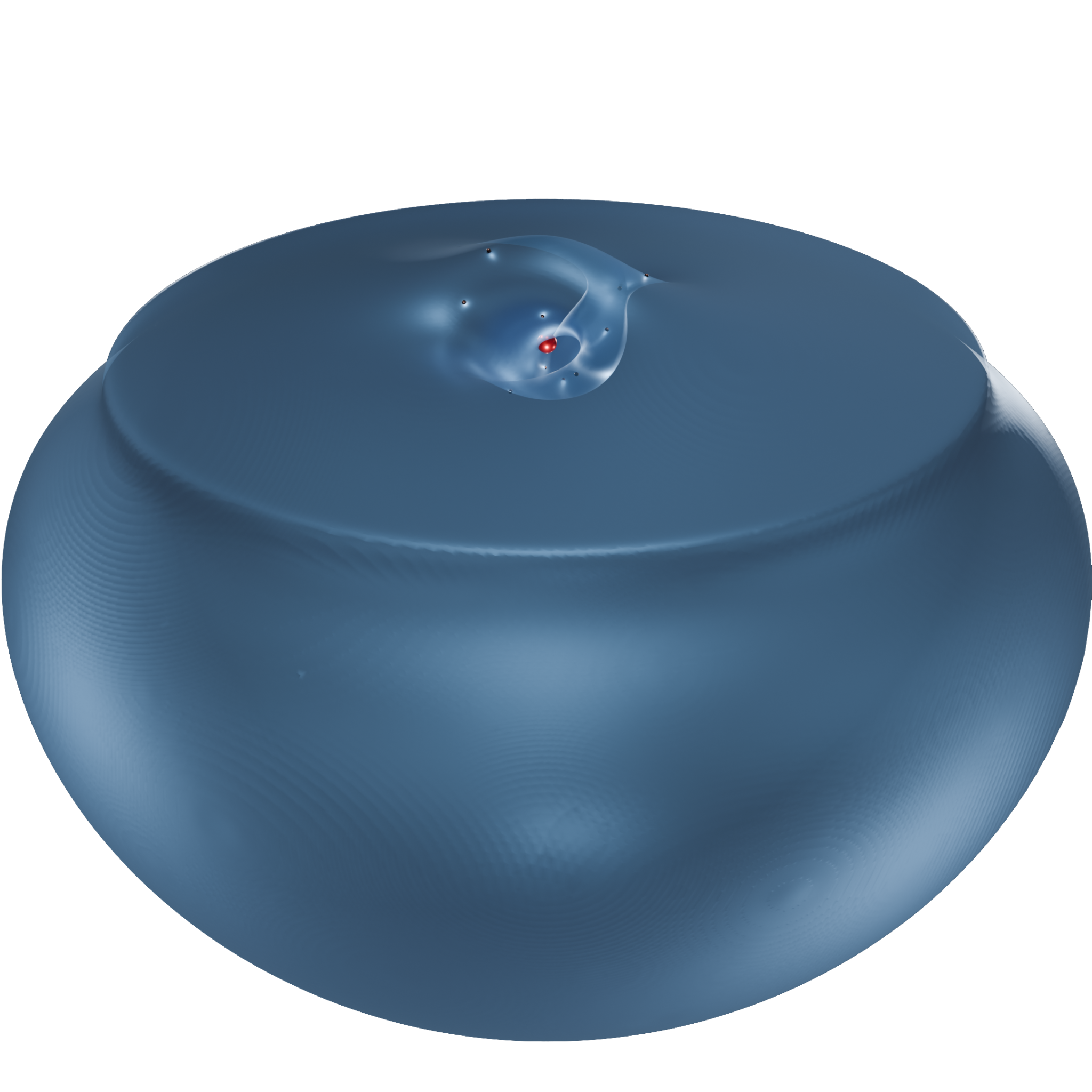}
    \caption{Nodal surface for the small-core Tl ccECP.}
    \label{fig:small_Tl_node}
\end{subfigure}
\caption{
Three-dimensional spatial projections of the single-determinant Hartree--Fock nodal hypersurface for the Tl atom. \textbf{(a)} Large-core variant ($Z_{\text{eff}}=3$, [$\text{[Xe]}4f^{14}5d^{10}$]), displaying a smooth, continuously connected valence nodal boundary. \textbf{(b)} Medium-core variant ($Z_{\text{eff}}=13$, [$\text{[Xe]}4f^{14}$]), illustrating a disconnected toroidal nodal domain near the core. \textbf{(c)} Small-core variant ($Z_{\text{eff}}=21$, [$\text{[Kr]}4d^{10}4f^{14}$]), demonstrating the restoration of a continuously connected nodal topology, characterized by compact spatial entanglement near the nucleus. The blue isosurface represents the topological manifold where the single-determinant many-body wavefunction evaluates to zero. To visualize this $(3N-1)$-dimensional boundary in 3D space, $N-1$ spin-aligned electrons (bright black spheres) are fixed at highly probable VMC equilibrium configurations, while a single probe electron scans the spatial grid (central red sphere denotes the nucleus).}
\label{fig:Tl_ccecps_nodal_surface}
\end{figure*}

The core-valence boundary ($Z_{\text{eff}}$) directly governs both the energy partitioning ratio $\xi$ and the geometric complexity of the spatial nodes. As shown in Fig.~\ref{fig:xi_6th_row}, shifting from a large- or medium-core to a small-core formulation substantially reduces $\xi$. For Hg, expanding the explicitly correlated space from $Z_{\text{eff}}=12$ to $20$ reduces $\xi$ from $68.855(17)\%$ to $33.531(12)\%$. Similarly, in Tl, $\xi$ falls from $61.911(19)\%$ ($Z_{\text{eff}}=13$) to $32.752(10)\%$ ($Z_{\text{eff}}=21$). This shift demonstrates that while un-pseudizing deep $5s5p$ semi-core states increases absolute kinetic energy ($E_{\text{CBS}}^{\text{kin}}$), the attractive potential energy near the nucleus deepens at a disproportionately faster rate, altering the overall energy balance and driving $\xi$ downward as $Z_{\text{eff}}$ expands. To visualize how $Z_{\text{eff}}$ shapes this spatial nodal complexity, Fig.~\ref{fig:Tl_ccecps_nodal_surface} displays three-dimensional cross-sections of single-determinant Hartree–Fock nodal hypersurfaces for the Tl $Z_{\text{eff}} = 3$, $13$, and $21$ active spaces. These surfaces represent 3D slices of the high-dimensional nodal manifold obtained by fixing $N-1$ electron coordinates. The contrast in complexity from a smooth, diffuse valence boundary (Fig.~\ref{fig:large_Tl_node}) to highly compact, entangled topologies near the origin (Fig.~\ref{fig:small_Tl_node}) visually captures the spatial confinement and enhanced nodal non-linearities, where rapid spatial wavefunction oscillations drive the escalation in kinetic energy~\cite{kevin2014}. A detailed analysis of how these specific topological variations influence the fixed-node error in FNDMC is deferred to Section \ref{subsec:fn_locality_bias}.

Beyond individual core formulations, $\xi$, serves as a global metric for evaluating spatial confinement, wavefunction curvature, and energy partitioning across the periodic table. Based on the calculated values tabulated in Table~\ref{tab:compact_DMC_bias_Ratio_kinetic_corr_methods_SA} and the profiles in Fig.~\ref{fig:xi_all_current_ccecps}, a systematic reduction in $\xi$ is observed moving down the periodic table. While second-period elements possess highly localized valence spaces (peaking at $80.902(6)$\% for Ne), sixth-period elements (Cs--Rn) remain confined within a band between $25.8$\% and $37.6$\%, reaching a minimum of $25.8449(86)$\% in Rn. This scaling reflects the departure of relativistic semi-local ECP Hamiltonians from the Coulombic virial relation ($2\langle T \rangle = -\langle V \rangle$, $\xi = 100\%$), as non-local projection operators and core node removal reduce kinetic energy near the origin~\cite{Ihm1979,Pickett1989,Bitter2010,Schwerdtfeger2011}.

\begin{table}[!htbp]
\setlength{\tabcolsep}{4pt}
\centering
\small
\caption{Extrapolated CBS expectation values of the valence kinetic energy ($E_{\text{CBS}}^{\text{kin}}$) for the sixth-period elements (Cs--Rn). Kinetic energies are evaluated from correlated many-body wavefunctions using CISD as a baseline. The methodology utilized to estimate the CBS limit is detailed in Sec. \ref{subsec:exact_ref_energy}. Statistical uncertainties are enclosed in parentheses. Additional tables containing all finite-basis $E_{nZ}$ values are provided in the Supplementary Information.}
\label{tab:compact_Kinetic_Energy_corr_methods_SA}
\begin{tabular}{lcccccc}
\hline
\textbf{Method} & \textbf{Atom} & \textbf{$\mathbf{Z_{\text{eff}}}$} & $E_{\rm CBS}^{\rm kin}$ (Ha) & \textbf{Atom} & \textbf{$\mathbf{Z_{\text{eff}}}$} & $E_{\rm CBS}^{\rm kin}$ (Ha) \\
\hline
CISD & \multirow{1}{*}{\textbf{Cs}} & \multirow{1}{*}{\textbf{9}} & 5.6704(11) & \multirow{1}{*}{\textbf{Ba}} & \multirow{1}{*}{\textbf{10}} & 7.0524(11) \\
\\
CISD & \multirow{1}{*}{\textbf{Hf}} & \multirow{1}{*}{\textbf{12}} & 16.7399(52) & \multirow{1}{*}{\textbf{Ta}} & \multirow{1}{*}{\textbf{13}} & 19.5768(63) \\
\\
CISD & \multirow{1}{*}{\textbf{W}} & \multirow{1}{*}{\textbf{14}} & 23.1116(69) & \multirow{1}{*}{\textbf{Re}} & \multirow{1}{*}{\textbf{15}} & 27.0789(83) \\
\\
CISD & \multirow{1}{*}{\textbf{Os}} & \multirow{1}{*}{\textbf{16}} & 31.193(11) & \multirow{1}{*}{\textbf{Ir}} & \multirow{1}{*}{\textbf{17}} & 35.234(12) \\
\\
CISD & \multirow{1}{*}{\textbf{Pt}} & \multirow{1}{*}{\textbf{18}} & 40.925(16) & \multirow{1}{*}{\textbf{Au}} & \multirow{1}{*}{\textbf{19}} & 46.311(17) \\
\\
CISD & \multirow{1}{*}{\textbf{Hg}} & \multirow{1}{*}{\textbf{12}} & 29.2787(73) & \multirow{1}{*}{\textbf{Hg}} & \multirow{1}{*}{\textbf{20}} & 51.447(18) \\
\\
CISD & \multirow{1}{*}{\textbf{Tl}} & \multirow{1}{*}{\textbf{13}} & 31.0484(94) & \multirow{1}{*}{\textbf{Tl}} & \multirow{1}{*}{\textbf{21}} & 56.467(17) \\
\\
CISD & \multirow{2}{*}{\textbf{Tl}} & \multirow{2}{*}{\textbf{3}} & 0.76139(50) & \multirow{2}{*}{\textbf{Pb}} & \multirow{2}{*}{\textbf{4}} & 1.13165(19) \\
FCI & & & 0.76366(53) & & & 1.13561(21) \\
\\
CISD & \multirow{2}{*}{\textbf{Bi}} & \multirow{2}{*}{\textbf{5}} & 1.90815(51) & \multirow{2}{*}{\textbf{Po}} & \multirow{2}{*}{\textbf{6}} & 2.41329(76) \\
FCI & & & 1.91414(52) & & & 2.42342(76) \\
\\
CISD & \multirow{2}{*}{\textbf{At}} & \multirow{2}{*}{\textbf{7}} & 3.1478(12) & \multirow{2}{*}{\textbf{Rn}} & \multirow{2}{*}{\textbf{8}} & 3.8340(13) \\
FCI & & & 3.1633(12) & & & 3.8397(13) \\
\end{tabular}
\end{table}

\begin{figure}[!htbp]
\centering
\begin{subfigure}{0.49\textwidth}
    \includegraphics[width=\textwidth]{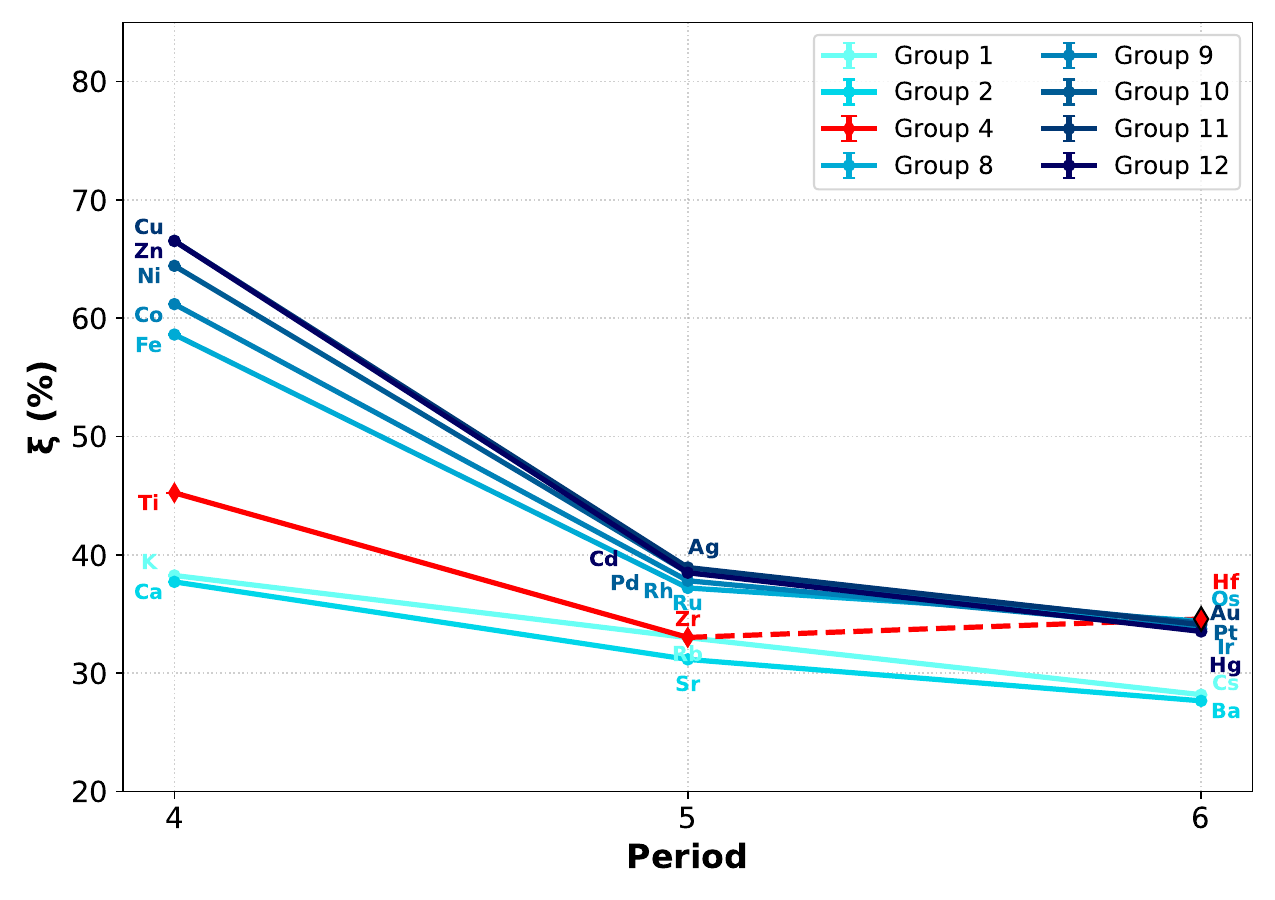}
    \caption{Group trends within the s- and d-block elements.}
    \label{fig:xi_trends_s_d_period}
\end{subfigure}\hfill
\begin{subfigure}{0.49\textwidth}
    \includegraphics[width=\textwidth]{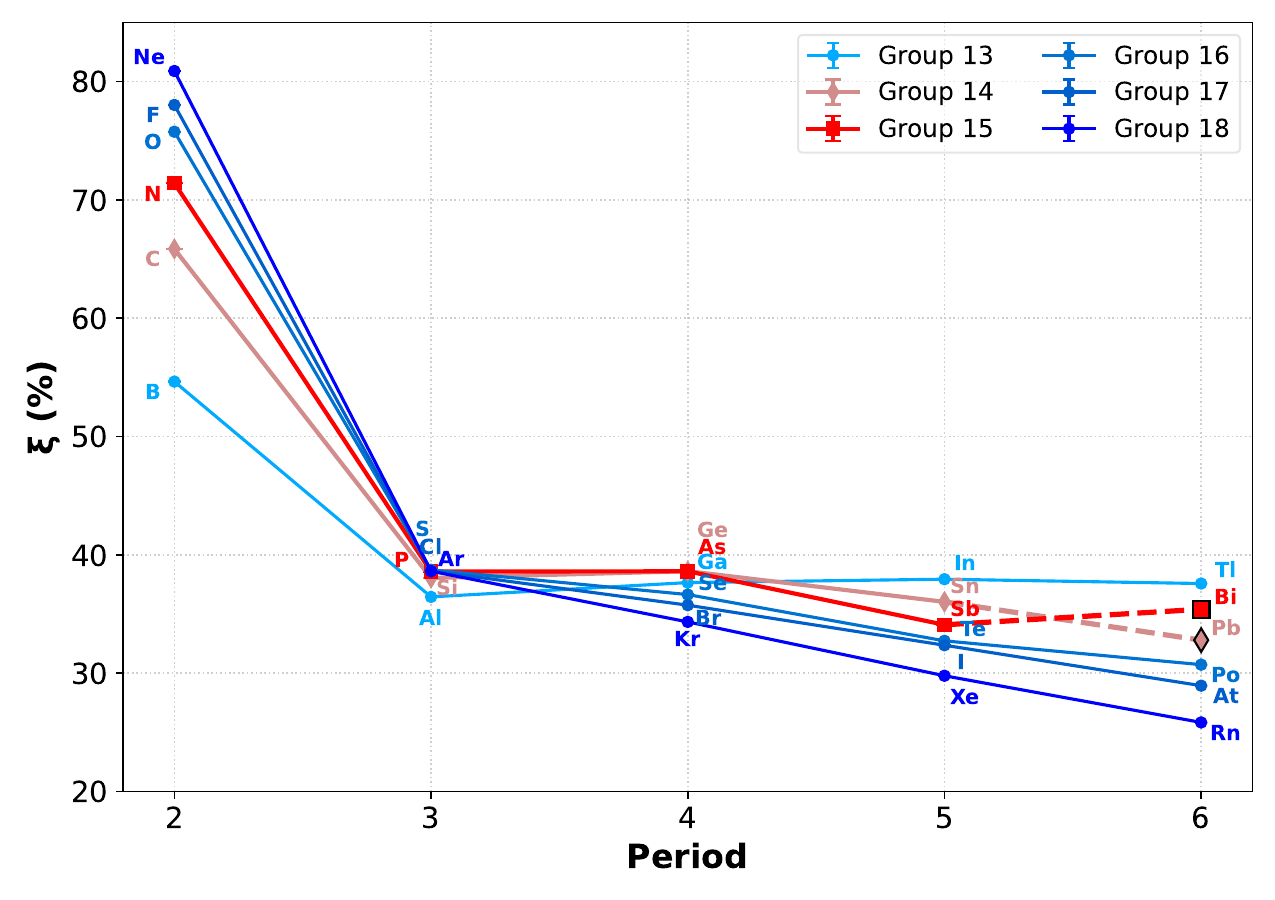}
    \caption{Group trends within the p-block elements.}
    \label{fig:xi_trends_p_period}
\end{subfigure}
\caption{Periodic trends in the valence kinetic-to-total energy ratio ($\xi$) as a function of the principal quantum number (period). Homologous elements are connected along group trajectories (groups 1--18) to illustrate the systematic attenuation of energy partitioning variation in heavier periods. Ratios for periods $2\text{--}5$ are evaluated from previously published ccECP data [Refs.~\cite{acc_engI,acc_engII}], while period $6$ incorporates the current work. \textbf{(a)} Attenuation of $\xi$ variation across s- and d-block elements (groups 1--12), where the flattening of the $5d$ series and the local maximum at Hf reflect the scalar relativistic contraction of the valence shell. \textbf{(b)} Trend reversal across the p-block elements (groups 13--18), where the general downward trend is driven by the spatial expansion and diffuse nature of the $6p$ outer orbitals.}
\label{fig:xi_trends_groups_period}
\end{figure}

The evolution of $\xi$ across transition-metal series highlights a quenching of kinetic energy variation in heavier periods (Fig.~\ref{fig:xi_trends_s_d_period}). The $3d$ series (period 4) exhibits a steep and monotonic increase from Sc ($41.336(9)$\%) to Cu ($66.547(5)$\%), reflecting the radial contraction and kinetic energy density escalation of the $3d$ shell. This trend is significantly attenuated in the $4d$ series and completely flattens in the $5d$ series (period 6). For the $5d$ elements, $\xi$ remains clustered near $34$\%. Within this $5d$ plateau, a distinct local maximum emerges at Hf ($34.584(11)$\%), breaking the expected downward group trend relative to its period-5 analog, Zr ($33.03(1)$\%)~\cite{acc_engII}. As Hf is the first transition metal immediately following the lanthanide series, it experiences the maximum relative impact of the lanthanide contraction~\cite{Kuchle1997,jiang2010}. Poor shielding provided by the $4f^{14}$ core exposes the valence electrons to a higher effective nuclear charge, driving a sudden radial contraction of the $5d$ and $6s$ orbitals that increases local wavefunction curvature and relative kinetic energy density compared to its lighter analog.

A qualitative trend reversal occurs in the $p$-block, as illustrated in Fig.~\ref{fig:xi_trends_p_period}. In period 2, increasing the nuclear charge tightens the compact $2p$ shell, which increases Pauli repulsion within a small volume and drives $\xi$ monotonically upward. In contrast, in period 6, the spatial extent of the $6p$ orbitals reverses this dynamic. Adding valence electrons increases electron-electron repulsion energy faster than nuclear confinement increases wavefunction curvature, driving a downward trend in $\xi$ toward Rn. An exception to this downward trend occurs at Bi ($35.4099(95)\%$), which likely reflects relativistic restructuring of the $6p$ valence density near the core boundary, a feature requiring further density analysis to fully disentangle from many-body correlation effects.

%%%===================================================%%%
%%%~~~~~~~~~~~~~~~~~~~~~~~~~~~~~~~~~~~~~~~~~~~~~~~~~~~%%%
%%%===================================================%%%

\begin{table*}[htbp!]
\setlength{\tabcolsep}{4pt}
\centering
\small
\caption{
Variational Monte Carlo (VMC), zero-timestep extrapolated diffusion Monte Carlo (DMC), and locality-error-corrected ($E_{\text{LEC}}$) total energies, absolute local energy variances ($\sigma^2$), and variance-to-energy ratios ($\sigma^2 / |E|$) with statistical uncertainties are enclosed in parentheses across sequential Jastrow optimization levels (no-J, J123). Calculations are performed under ccECP Hamiltonians utilizing the T-moves localization scheme. The $\sigma^2 / |E|$ ratio serves as an intensive, size-consistent metric of wavefunction optimization quality across varying core-valence sizes. Extrapolation parameters are discussed in Sec. \ref{subsec:qmc}. Individual timestep DMC total energies and DMC/VMC kinetic-energy data are provided in Supplementary Table 8--9. Visualizations of the DMC timestep extrapolations and pseudopotential locality error estimations are provided in Supplementary Figures 1--4.}
\label{tab:VMC_Total_Energy_noJ_J123}
\begin{tabular}{c*{5}{c}*{5}{c}}
\hline
\textbf{Method} & \textbf{Element} & \textbf{$\mathbf{Z_{\text{eff}}}$} & Energy (Ha) & Variance (Ha$^2$) & $|\sigma^2/E|$ (Ha) & \textbf{Element} & \textbf{$\mathbf{Z_{\text{eff}}}$}  & Energy (Ha) & Variance (Ha$^2$) & $|\sigma^2/E|$ (Ha) \\
\hline\hline
\textbf{VMC-noJ} & \multirow{4}{*}{\textbf{Cs}} & \multirow{4}{*}{\textbf{9}} & -19.86441(35) & 1.2822(66) & 0.0645 & \multirow{4}{*}{\textbf{Ba}} & \multirow{4}{*}{\textbf{10}} & -25.17316(38) & 1.631(31) & 0.0648 \\
\textbf{VMC-J123} & & & -20.098892(79) & 0.1104(36) & 0.0055 & & & -25.45551(11) & 0.1862(14) & 0.0073 \\
\textbf{DMC ($\tau \to 0$)} & & & -20.107673(82) & 0.10605(14) & 0.0053 & & & -25.47176(34) & 0.1846(18) & 0.0072 \\
\textbf{$\mathbf{E_{\text{LEC}}}$} & & & -20.10912(57) & -- & -- & & & -25.4744(19) & -- & -- \\
\\
\textbf{VMC-noJ} & \multirow{4}{*}{\textbf{Hf}} & \multirow{4}{*}{\textbf{12}} & -48.01184(53) & 3.882(56) & 0.0809 & \multirow{4}{*}{\textbf{Ta}} & \multirow{4}{*}{\textbf{13}} & -56.85849(89) & 4.66(10) & 0.0820 \\
\textbf{VMC-J123} & & & -48.36118(16) & 0.3196(22) & 0.0066 & & & -57.24489(18) & 0.43072(93) & 0.0075 \\
\textbf{DMC ($\tau \to 0$)} & & & -48.38573(15) & 0.318661(18) & 0.0066 & & & -57.27025(45) & 0.4378(39) & 0.0076 \\
\textbf{$\mathbf{E_{\text{LEC}}}$} & & & -48.39207(46) & -- & -- & & & -57.2781(15) & -- & -- \\
\\
\textbf{VMC-noJ} & \multirow{4}{*}{\textbf{W}} & \multirow{4}{*}{\textbf{14}} & -66.69734(59) & 5.376(33) & 0.0806 & \multirow{4}{*}{\textbf{Re}} & \multirow{4}{*}{\textbf{15}} & -77.77497(98) & 6.274(41) & 0.0807 \\
\textbf{VMC-J123} & & & -67.13007(19) & 0.4677(36) & 0.0070 & & & -78.26410(18) & 0.6312(33) & 0.0081 \\
\textbf{DMC ($\tau \to 0$)} & & & -67.15716(43) & 0.4637(22) & 0.0069 & & & -78.291858(98) & 0.6315(27) & 0.0081 \\
\textbf{$\mathbf{E_{\text{LEC}}}$} & & & -67.1643(17) & -- & -- & & & -78.29699(28) & -- & -- \\
\\
\textbf{VMC-noJ} & \multirow{4}{*}{\textbf{Os}} & \multirow{4}{*}{\textbf{16}} & -90.02452(84) & 7.592(37) & 0.0843 & \multirow{4}{*}{\textbf{Ir}} & \multirow{4}{*}{\textbf{17}} & -103.6668(12) & 9.063(55) & 0.0874 \\
\textbf{VMC-J123} & & & -90.57743(31) & 0.822(16) & 0.0091 & & & -104.28441(28) & 0.8856(33) & 0.0085 \\
\textbf{DMC ($\tau \to 0$)} & & & -90.61079(89) & 0.8119(49) & 0.0090 & & & -104.31935(35) & 0.8850(12) & 0.0085 \\
\textbf{$\mathbf{E_{\text{LEC}}}$} & & & -90.6149(11) & -- & -- & & & -104.3258(12) & -- & -- \\
\\
\textbf{VMC-noJ} & \multirow{4}{*}{\textbf{Pt}} & \multirow{4}{*}{\textbf{18}} & -118.7591(11) & 11.11(15) & 0.0936 & \multirow{4}{*}{\textbf{Au}} & \multirow{4}{*}{\textbf{19}} & -135.00657(58) & 12.826(27) & 0.0950 \\
\textbf{VMC-J123} & & & -119.43816(32) & 1.1801(88) & 0.0099 & & & -135.74901(26) & 1.3260(53) & 0.0098 \\
\textbf{DMC ($\tau \to 0$)} & & & -119.47452(19) & 1.1786(29) & 0.0099 & & & -135.78439(60) & 1.3288(32) & 0.0098 \\
\textbf{$\mathbf{E_{\text{LEC}}}$} & & & -119.48033(32) & -- & -- & & & -135.7903(12) & -- & -- \\
\\
\textbf{VMC-noJ} & \multirow{4}{*}{\textbf{Hg}} & \multirow{4}{*}{\textbf{12}} & -41.82135(69) & 6.557(32) & 0.1568 & \multirow{4}{*}{\textbf{Hg}} & \multirow{4}{*}{\textbf{20}} & -152.5314(15) & 15.36(51) & 0.1007 \\
\textbf{VMC-J123} & & & -42.44429(25) & 0.7864(35) & 0.0185 & & & -153.35024(29) & 1.1890(40) & 0.0078 \\
\textbf{DMC ($\tau \to 0$)} & & & -42.479126(56) & 0.7862(10) & 0.0185 & & & -153.38871(25) & 1.18786(70) & 0.0077 \\
\textbf{$\mathbf{E_{\text{LEC}}}$} & & & -42.4864(10) & -- & -- & & & -153.39372(25) & -- & -- \\
\\
\textbf{VMC-noJ} & \multirow{4}{*}{\textbf{Tl}} & \multirow{4}{*}{\textbf{13}} & -49.44037(68) & 7.054(15) & 0.1427 & \multirow{4}{*}{\textbf{Tl}} & \multirow{4}{*}{\textbf{21}} & -171.4832(11) & 16.241(83) & 0.0947 \\
\textbf{VMC-J123} & & & -50.08249(23) & 0.8306(27) & 0.0166 & & & -172.33448(33) & 1.3775(97) & 0.0080 \\
\textbf{DMC ($\tau \to 0$)} & & & -50.11947(88) & 0.8335(22) & 0.0166 & & & -172.37353(43) & 1.3672(19) & 0.0079 \\
\textbf{$\mathbf{E_{\text{LEC}}}$} & & & -50.1308(42) & -- & -- & & & -172.37866(37) & -- & -- \\
\\
\textbf{VMC-noJ} & \multirow{4}{*}{\textbf{Tl}} & \multirow{4}{*}{\textbf{3}} & -1.98799(10) & 0.0970(17) & 0.0488 & \multirow{4}{*}{\textbf{Pb}} & \multirow{4}{*}{\textbf{4}} & -3.40232(20) & 0.1611(22) & 0.0474 \\
\textbf{VMC-J123} & & & -2.029225(20) & 0.007699(41) & 0.0038 & & & -3.457340(35) & 0.014802(42) & 0.0043 \\
\textbf{DMC ($\tau \to 0$)} & & & -2.031467(94) & 0.007474(50) & 0.0037 & & & -3.460040(94) & 0.01476(13) & 0.0043 \\
\textbf{$\mathbf{E_{\text{LEC}}}$} & & & -2.031889(20) & -- & -- & & & -3.46122(30) & -- & -- \\
\\
\textbf{VMC-noJ} & \multirow{4}{*}{\textbf{Bi}} & \multirow{4}{*}{\textbf{5}} & -5.32785(22) & 0.2550(48) & 0.0479 & \multirow{4}{*}{\textbf{Po}} & \multirow{4}{*}{\textbf{6}} & -7.76476(18) & 0.4273(80) & 0.0550 \\
\textbf{VMC-J123} & & & -5.398053(65) & 0.0428(63) & 0.0079 & & & -7.874996(59) & 0.03731(20) & 0.0047 \\
\textbf{DMC ($\tau \to 0$)} & & & -5.40237(20) & 0.03108(82) & 0.0058 & & & -7.882149(91) & 0.037310(90) & 0.0047 \\
\textbf{$\mathbf{E_{\text{LEC}}}$} & & & -5.40446(59) & -- & -- & & & -7.88363(16) & -- & -- \\
\\
\textbf{VMC-noJ} & \multirow{4}{*}{\textbf{At}} & \multirow{4}{*}{\textbf{7}} & -10.75720(21) & 0.6493(24) & 0.0604 & \multirow{4}{*}{\textbf{Rn}} & \multirow{4}{*}{\textbf{8}} & -14.64370(31) & 0.914(11) & 0.0624 \\
\textbf{VMC-J123} & & & -10.909545(57) & 0.05658(39) & 0.0052 & & & -14.836539(73) & 0.08593(30) & 0.0058 \\
\textbf{DMC ($\tau \to 0$)} & & & -10.91823(12) & 0.05649(35) & 0.0052 & & & -14.845106(33) & 0.08632(43) & 0.0058 \\
\textbf{$\mathbf{E_{\text{LEC}}}$} & & & -10.91995(18) & -- & -- & & & -14.84757(41) & -- & -- \\
\end{tabular}
\end{table*}

% \clearpage
% \newpage
\subsection{Variational and Diffusion Monte Carlo Energies}
\label{subsec:vmc_dmc_energies}

Table~\ref{tab:VMC_Total_Energy_noJ_J123} reports VMC and extrapolated zero-timestep DMC energies, local energy variances ($\sigma^2$), and variance-to-energy ratios ($|\sigma^2/E|$). Optimizing three-body Jastrow factors (VMC-J123) reduces local energy variance $\sigma^2$ by more than an order of magnitude relative to bare Slater determinants (VMC-noJ). This reduction demonstrates that parameterized Jastrow factors efficiently recover valence dynamic correlation.

For most Cs--Rn elements, the $\sigma^2 / |E|$ ratio remains below $0.01~\text{Ha}$, demonstrating well-controlled local energy noise across the series. However, absolute variance increases sharply with $Z_{\text{eff}}$. In thallium, expanding the explicitly correlated space from $Z_{\text{eff}}=3$ to $Z_{\text{eff}}=21$ increases the variance from $0.0077\text{ Ha}^2$ to $1.3775\text{ Ha}^2$. As illustrated by the one-particle nodal cross-sections in Fig.~\ref{fig:Tl_ccecps_nodal_surface}, un-pseudizing the $5s5p$ and $5d$ shells introduces dense, highly localized nodal features near the nucleus. Despite the inherent high variance of these small-core configurations, systematic timestep extrapolations ($\tau \to 0$) and sufficiently large walker populations still yield sub-millihartree DMC statistical precision ($\pm 0.43$~mHa for Tl, $Z_{\text{eff}}=21$). Close agreement between projected FNDMC energies and CBS CCSDT(Q) reference limits provides initial validation of the single-reference trial wavefunctions; however, rigorously disentangling nodal fidelity from pseudopotential localization errors requires the detailed partitioning analysis in Section \ref{subsec:fn_locality_bias}. 

%%%===================================================%%%
%%%~~~~~~~~~~~~~~~~~~~~~~~~~~~~~~~~~~~~~~~~~~~~~~~~~~~%%%
%%%===================================================%%%

% \newpage
% \clearpage
% \newpage
\subsection{Assessment of Fixed-Node and Locality Biases}
\label{subsec:fn_locality_bias}

\begin{figure}[!htbp]
\centering
\begin{subfigure}{0.49\textwidth}
    \includegraphics[width=\textwidth]{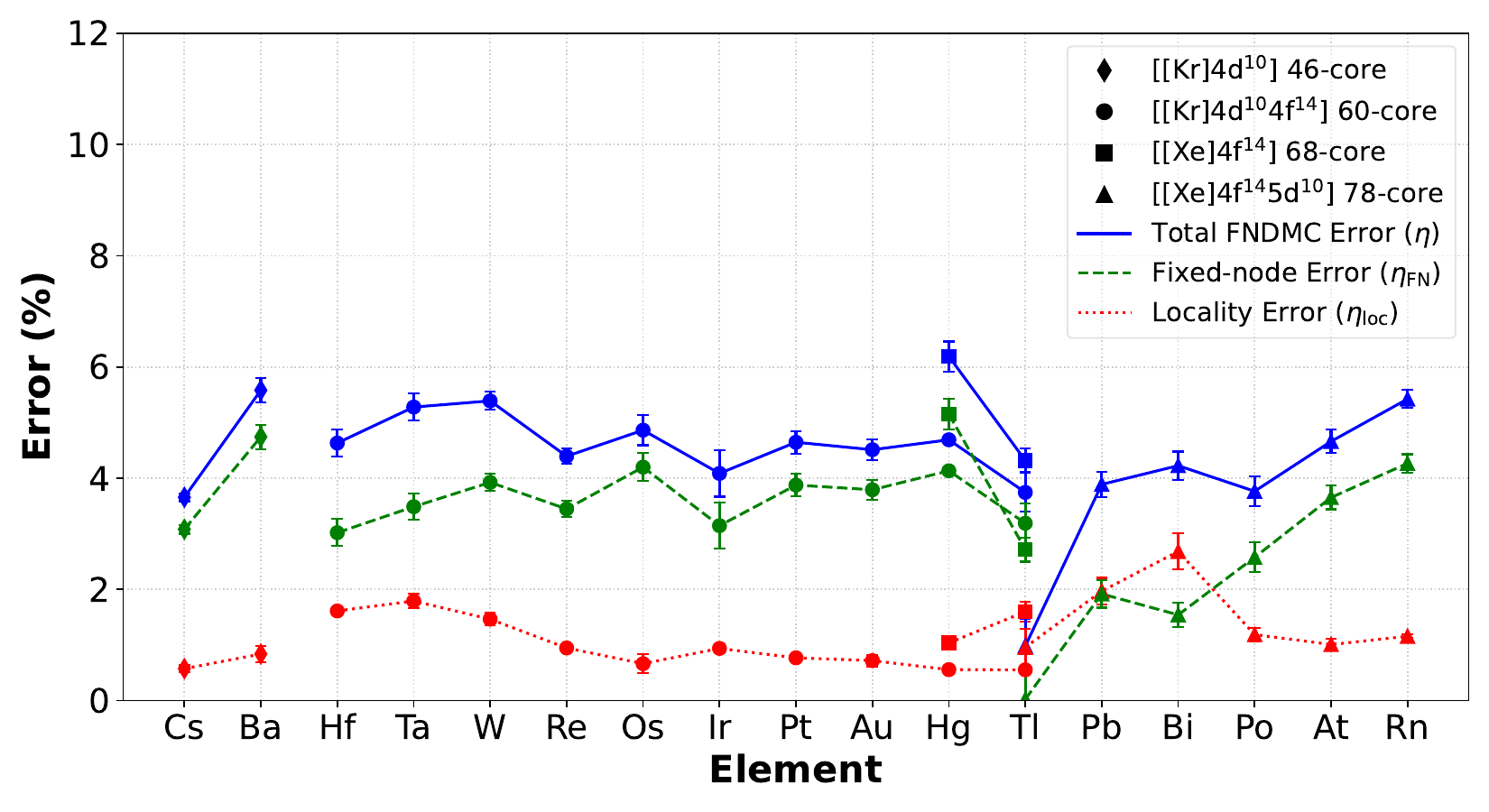}
    \caption{Relative error decomposition for the Cs--Rn series.}
    \label{fig:eta_6th_row}
\end{subfigure}\hfill
\begin{subfigure}{0.49\textwidth}
    \includegraphics[width=\textwidth]{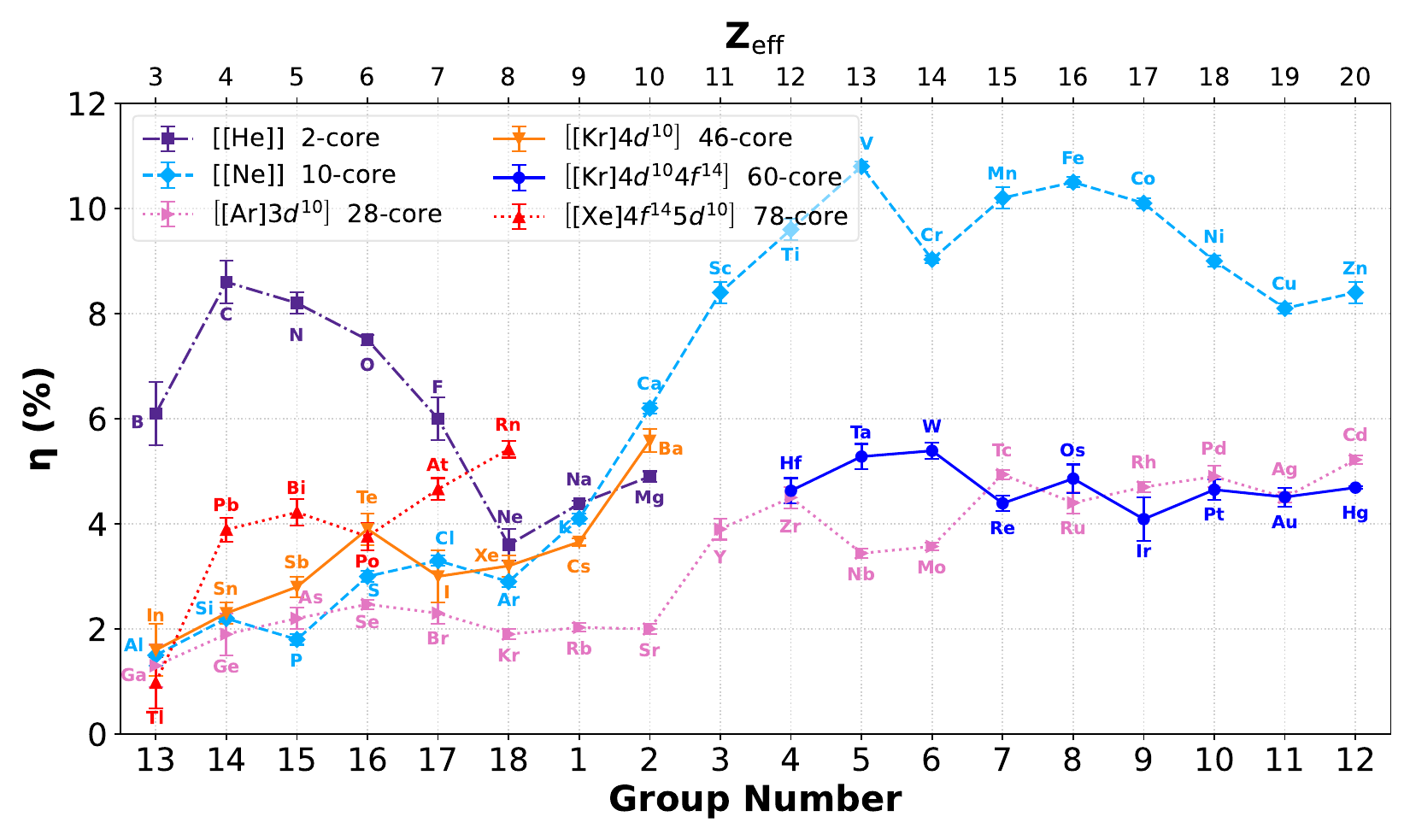}
    \caption{Extended periodic comparison of relative combined FNDMC bias $\eta$.}
    \label{fig:eta_all_current_ccecps}
\end{subfigure}
\caption{Relative combined FNDMC bias ($\eta$) and its partitioned components across the periodic table. \textbf{(a)} Partitioning of the relative total bias ($\eta$, Eq. \ref{eq:eta}) into relative fixed-node ($\eta_{\text{FN}}$, Eq. \ref{eq:eta_FN}) and relative locality ($\eta_{\text{loc}}$, Eq. \ref{eq:eta_loc}) errors for the Cs--Rn series as a function of the effective valence charge ($Z_{\text{eff}}$). This decomposition utilizes the zero-timestep extrapolated T-moves scheme and the locality-error-corrected regression limit. \textbf{(b)} Extended periodic-table-wide comparison of the relative combined bias $\eta$, integrating the current Cs--Rn results with previously published ccECPs [Refs.~\cite{acc_engI,acc_engII}], ordered by effective core charge ($Z_{\text{eff}}$), periodic group, and core size. All metrics are normalized against the CBS valence correlation energy, ensuring a size-consistent comparison across diverse valence space dimensions. Numerical values are provided in Tables~\ref{tab:compact_DMC_bias_Ratio_kinetic_corr_methods_SA} and~\ref{tab:DMC_bias_loc_fn_SA}.}
\label{fig:eta_core_size_plots}
\end{figure}

\begin{figure}[!htbp]
\centering
\begin{subfigure}{0.49\textwidth}
    \includegraphics[width=\textwidth]{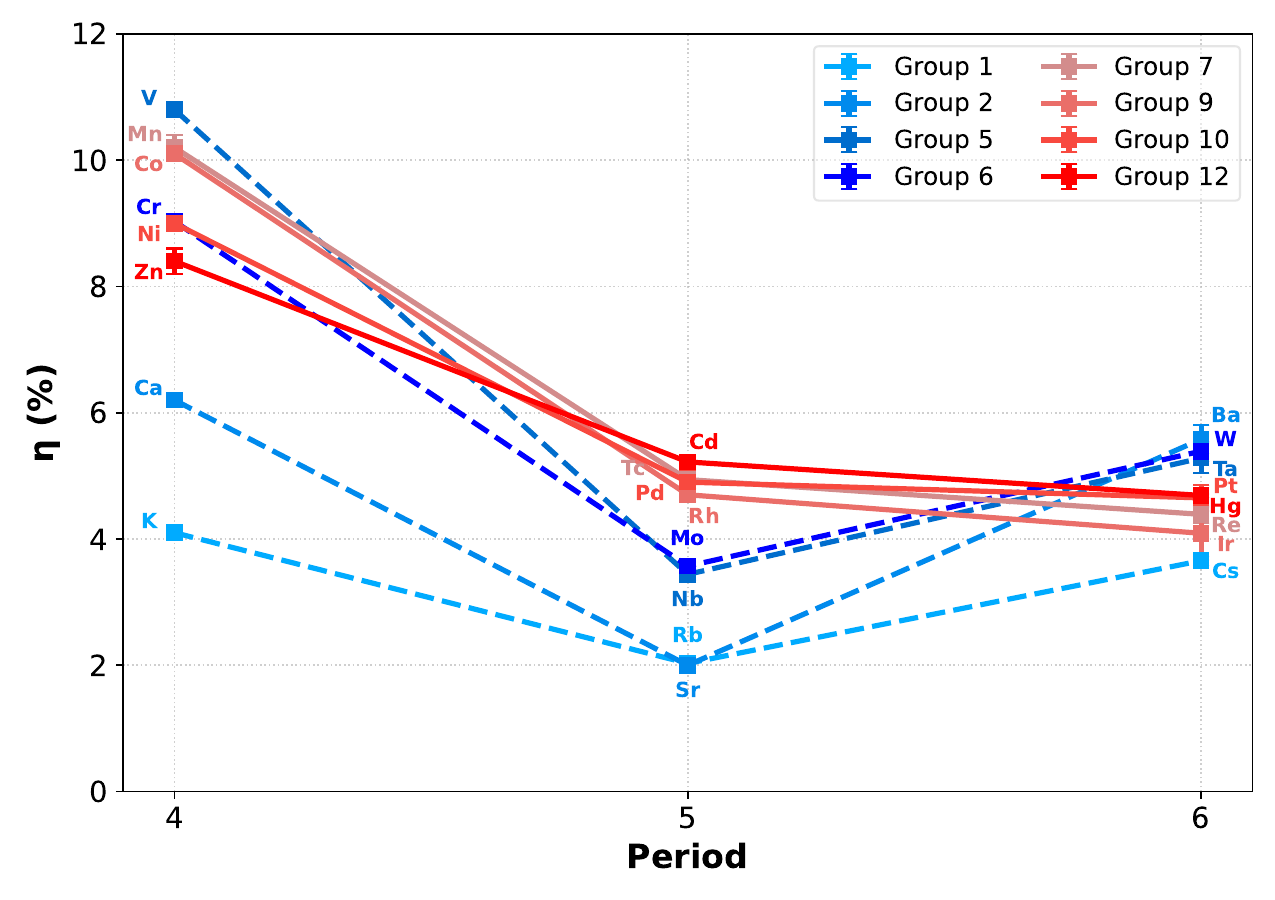}
    \caption{Group trends within the s- and d-block elements}
    \label{fig:eta_trends_s_d_period}
\end{subfigure}\hfill
\begin{subfigure}{0.49\textwidth}
    \includegraphics[width=\textwidth]{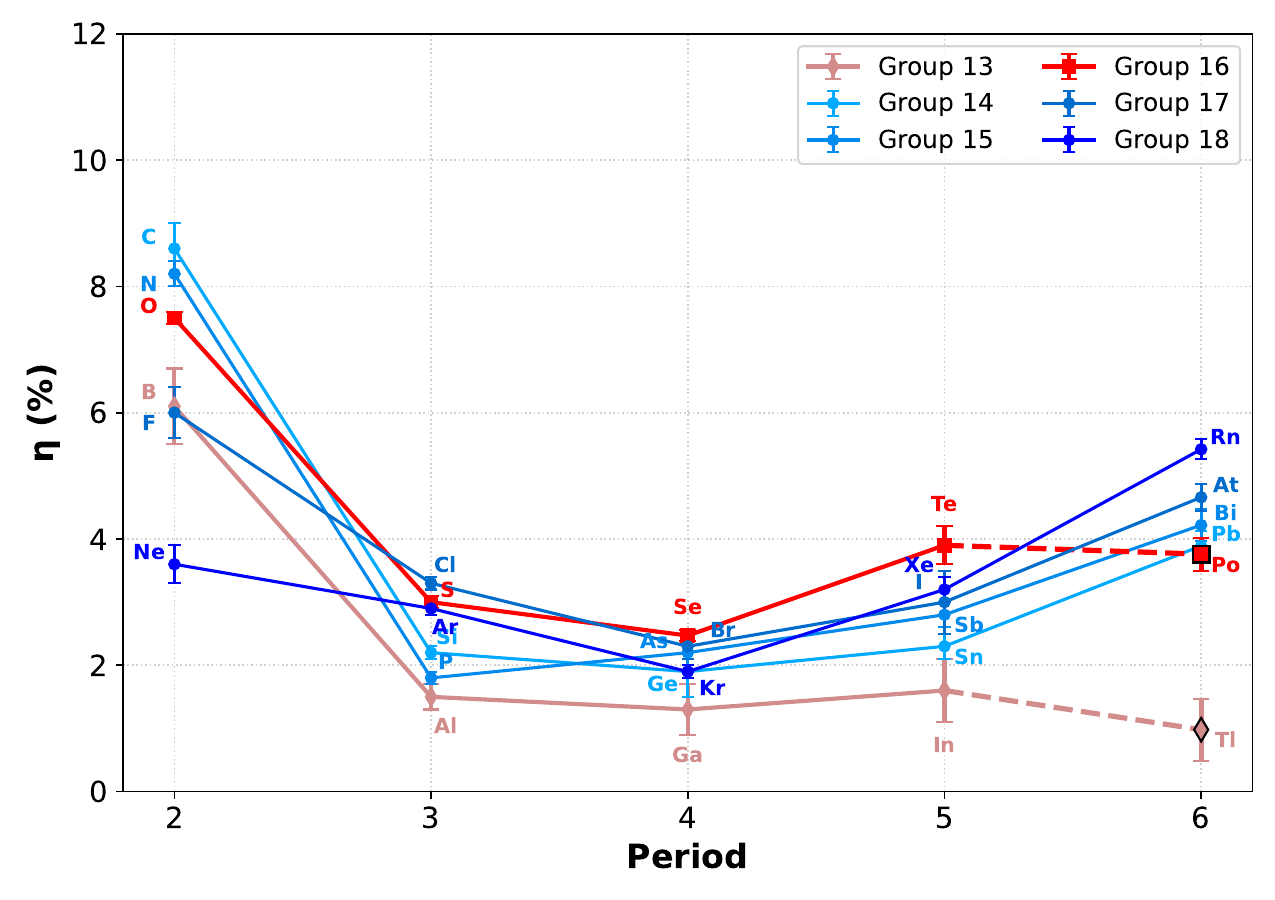}
    \caption{Group trends within the p-block elements.}
    \label{fig:eta_trends_p_period}
\end{subfigure}
\caption{Periodic trends in the fixed-node and locality bias ($\eta$) as a function of the respective principal quantum number and period. Homologous elements are connected along group trajectories (groups 1--18). Data points for periods $n=2\text{--}5$ are obtained from previously published ccECP studies [Refs.~\cite{acc_engI,acc_engII}], while period $n=6$ incorporates the current work. \textbf{(a)} Trends within the s- and d-block elements (groups 1–12), where the dramatic drop in $\eta$ from the $3d$ till $5d$ series illustrates the relaxation of nodal complexity driven by core orthogonality constraints. \textbf{(b)} Trends within the p-block elements (groups 13–18), where the steep rise toward the closed-shell limit reflects the performance of the single-reference ansatz under high-density.
}
\label{fig:eta_trends_groups_period}
\end{figure}

Table~\ref{tab:compact_DMC_bias_Ratio_kinetic_corr_methods_SA} reports the relative combined bias, $\eta$ (Eq.~\ref{eq:eta}), which represents the percentage of the CBS valence correlation energy not recovered by FNDMC using a single-reference Slater-Jastrow trial wavefunction. We utilize this normalized metric to systematically compare the combined fixed-node and localization errors across the Cs--Rn series, plotting our results alongside previous work~\cite{acc_engI,acc_engII} for lighter analogs in Figs.~\ref{fig:eta_core_size_plots}--\ref{fig:eta_trends_groups_period}.

\begin{table}[htbp!]
\renewcommand{\arraystretch}{1.0}
\setlength{\tabcolsep}{6pt}
\centering
\small
\caption{Calculated kinetic-to-total energy partitioning ratio ($\xi$, Eq. \ref{eq:xi}) and relative combined FNDMC bias ($\eta$, Eq. \ref{eq:eta}) for the sixth-period elements (Cs--Rn) expressed as percentages. The exact baseline properties are evaluated at the CCSDT(Q)/CBS limit. The FNDMC calculations utilize single-reference Slater-Jastrow trial wavefunctions with three-body Jastrow factors  and are extrapolated to the zero-timestep limit under the T-moves localization scheme. Statistical uncertainties are provided in parentheses. Crucially, $\xi$ is an exact physical property of the valence-only Hamiltonian, whereas $\eta$ is an algorithm-dependent metric of FNDMC bias normalized against the CBS correlation energy.}
\label{tab:compact_DMC_bias_Ratio_kinetic_corr_methods_SA}
\begin{tabular}{ccccc}
\hline
\textbf{Atom} & \textbf{$\mathbf{Z_{\text{eff}}}$} & $\epsilon$~(mHa) & $\eta$~(\%) & $\xi$~(\%) \\
\hline
\textbf{Cs} & \textbf{9} & 9.22(18) & 3.656(70) & 28.1874(52) \\
\textbf{Ba} & \textbf{10} & 17.68(71) & 5.58(22) & 27.6679(44) \\
\textbf{Hf} & \textbf{12} & 18.23(95) & 4.63(24) & 34.584(11) \\
\textbf{Ta} & \textbf{13} & 23.1(1.1) & 5.28(24) & 34.169(11) \\
\textbf{W} & \textbf{14} & 26.39(79) & 5.39(16) & 34.401(10) \\
\textbf{Re} & \textbf{15} & 23.79(76) & 4.39(14) & 34.577(11) \\
\textbf{Os} & \textbf{16} & 30.2(1.7) & 4.86(27) & 34.414(12) \\
\textbf{Ir} & \textbf{17} & 27.9(2.8) & 4.09(42) & 33.766(12) \\
\textbf{Pt} & \textbf{18} & 35.1(1.5) & 4.65(20) & 34.244(14) \\
\textbf{Au} & \textbf{19} & 36.8(1.5) & 4.51(18) & 34.097(13) \\
\textbf{Hg} & \textbf{12} & 43.3(1.9) & 6.19(27) & 68.855(17) \\
\textbf{Hg} & \textbf{20} & 42.19(26) & 4.689(29) & 33.531(12) \\
\textbf{Tl} & \textbf{13} & 30.7(1.5) & 4.32(21) & 61.911(19) \\
\textbf{Tl} & \textbf{21} & 34.7(3.2) & 3.74(35) & 32.752(10) \\
\textbf{Tl} & \textbf{3} & 0.43(21) & 0.98(49) & 37.584(26) \\
\textbf{Pb} & \textbf{4} & 2.34(14) & 3.89(23) & 32.7985(61) \\
\textbf{Bi} & \textbf{5} & 3.28(20) & 4.22(25) & 35.4099(95) \\
\textbf{Po} & \textbf{6} & 4.69(33) & 3.76(26) & 30.7274(97) \\
\textbf{At} & \textbf{7} & 7.95(36) & 4.66(21) & 28.951(11) \\
\textbf{Rn} & \textbf{8} & 11.54(34) & 5.42(16) & 25.8449(86) \\
\end{tabular}
\end{table}

At the onset of period 6, we observe a significant escalation in bias from $\text{Cs} = 3.656(70)\%$ to $\text{Ba} = 5.58(22)\%$. This systematic increase mirrors the fourth-period transition from $\text{K} = 4.1(1)\% \to \text{Ca} = 6.2(1)\%$, and departs from the anomalous flatness observed in the fifth-period (where Rb and Sr both produce $\eta \approx 2.0\%$). The $5d$ transition metal series (Hf--Hg) exhibits a combined bias $\eta$ that remains bounded between 4.09(42)\% (Ir) and 5.39(16)\% (W), indicating that the single-reference nodal surface maintains a stable level of fidelity despite the significant increase in the effective nuclear charge $Z_{\text{eff}}$. The relative bias $\eta$ decreases from the compact $3d$ series to the diffuse $4d$ and $5d$ series (Fig.~\ref{fig:eta_trends_s_d_period}). Unlike the highly localized $3d$ shell that lacks core states of the same angular momentum and exhibits steep local density gradients~\cite{mitas2025}, the $4d$ and $5d$ shells must maintain orthogonality against the inner core states. This requirement forces the valence orbitals to become more diffuse, smoothing local density gradients and presumably reducing the topological complexity of the nodal manifold~\cite{kevin2014}. Notably, the $5d$ series lacks the characteristic decrease in bias near the half-filled shell observed in the $3d$ (Cr)~\cite{acc_engI} and $4d$ (Nb, Mo)~\cite{acc_engII} series. In the $5d$ row, the competing relativistic expansion of the $5d$ shell and the contraction of the $6s$ shell maintain an exceptionally polarizable core-valence boundary. This prevents any sharp, exchange-driven nodal simplification, keeping $\eta$ remarkably flat and bounded near 4.5\% from Hf to Hg.

The $6p$ main-group series (Tl–-Rn), evaluated using a large 78-electron $[Xe]4f^{14}5d^{10}$ core, exhibits $\eta$ values ranging from $0.98(49)$\% in Tl to $5.42(16)$\% in Rn. Unlike the relatively flat profiles observed in the $3p$--$5p$ blocks, the $6p$ series exhibits a rapid escalation toward the closed-shell limit. The intermediate elements Pb, Bi, and Po form a statistically indistinguishable plateau with overlapping uncertainties. Beyond this plateau, the strong increase of $\eta$ for At and Rn mirrors the density-driven correlation errors documented in AE studies of heavy noble gases. In this dense, closed-shell regime, multi-configurational contributions restructure the exact nodal manifold beyond what a single-determinant trial wavefunction can accommodate, sharply increasing the unrecovered correlation energy~\cite{mitas2025}.
\begin{table}[!htbp]
\setlength{\tabcolsep}{6pt}
\centering
\small
\caption{Absolute and relative fixed-node ($\epsilon_{\text{FN}}$, $\eta_{\text{FN}}$) and locality ($\epsilon_{\text{loc}}$, $\eta_{\text{loc}}$) errors for the sixth-period elements (Cs--Rn). Absolute errors are reported in mHa, and relative errors are expressed as percentages of the CBS valence correlation energy (Eq.\ref{eq:eta_loc}--\ref{eq:eta_FN}). The errors are partitioned by extracting the locality-error-corrected limit from the linear fit of DMC and VMC energies as described in Sec.\ref{subsec:qmc}. This separation isolates the intrinsic topological error of the single-reference nodal surface from the localization error introduced by the T-moves approximation.}
\label{tab:DMC_bias_loc_fn_SA}
\begin{tabular}{cccccc}
\hline
\textbf{Atom} & \textbf{$\mathbf{Z_{\text{eff}}}$} & $\epsilon_{\text{loc}}$~(mHa) & $\epsilon_{\text{FN}}$~(mHa) & $\eta_{\text{loc}}$ (\%) & $\eta_{\text{FN}}$ (\%) \\
\hline
\textbf{Cs} & \textbf{9} & 1.45(15) & 7.77(20) & 0.575(59) & 3.081(79) \\
\textbf{Ba} & \textbf{10} & 2.65(46) & 15.03(69) & 0.84(14) & 4.74(22) \\
\textbf{Hf} & \textbf{12} & 6.34(22) & 11.88(96) & 1.612(57) & 3.02(24) \\
\textbf{Ta} & \textbf{13} & 7.83(58) & 15.2(1.0) & 1.79(13) & 3.49(23) \\
\textbf{W} & \textbf{14} & 7.18(56) & 19.22(75) & 1.47(11) & 3.92(15) \\
\textbf{Re} & \textbf{15} & 5.13(17) & 18.66(77) & 0.947(31) & 3.45(14) \\
\textbf{Os} & \textbf{16} & 4.1(1.1) & 26.1(1.6) & 0.66(18) & 4.20(25) \\
\textbf{Ir} & \textbf{17} & 6.41(53) & 21.5(2.9) & 0.939(78) & 3.15(42) \\
\textbf{Pt} & \textbf{18} & 5.81(32) & 29.3(1.5) & 0.770(43) & 3.88(20) \\
\textbf{Au} & \textbf{19} & 5.87(83) & 30.9(1.5) & 0.72(10) & 3.79(18) \\
\textbf{Hg} & \textbf{12} & 7.26(19) & 36.0(1.9) & 1.037(28) & 5.15(27) \\
\textbf{Hg} & \textbf{20} & 5.01(47) & 37.17(40) & 0.557(52) & 4.132(45) \\
\textbf{Tl} & \textbf{13} & 11.4(1.3) & 19.3(1.6) & 1.60(18) & 2.72(22) \\
\textbf{Tl} & \textbf{21} & 5.14(70) & 29.5(3.2) & 0.555(76) & 3.19(35) \\
\textbf{Tl} & \textbf{3} & 0.42(15) & 0.01(22) & 0.96(33) & 0.02(51) \\
\textbf{Pb} & \textbf{4} & 1.18(14) & 1.15(15) & 1.97(24) & 1.92(25) \\
\textbf{Bi} & \textbf{5} & 2.09(25) & 1.20(17) & 2.68(33) & 1.54(21) \\
\textbf{Po} & \textbf{6} & 1.48(15) & 3.21(33) & 1.19(12) & 2.58(27) \\
\textbf{At} & \textbf{7} & 1.72(18) & 6.23(36) & 1.01(10) & 3.65(21) \\
\textbf{Rn} & \textbf{8} & 2.463(76) & 9.08(35) & 1.157(36) & 4.27(16) \\
\end{tabular}
\end{table}

Partitioning the total DMC bias (Fig.~\ref{fig:eta_6th_row} and Table~\ref{tab:DMC_bias_loc_fn_SA}) reveals that the fixed-node error ($\epsilon_{\mathrm{FN}}$) dominates the locality error ($\epsilon_{\mathrm{loc}}$) across most sixth-period elements, confirming nodal topology of single-determinant trial wavefunctions as the primary bottleneck. However, notable deviations from this trend occur in the $6p$ block. Bismuth ($Z_{\text{eff}}=5$) is the only case in the studied series where the locality contribution exceeds the fixed-node contribution ($\epsilon_{\mathrm{loc}} = 2.09(25)$ mHa vs $\epsilon_{\mathrm{FN}} = 1.20(17)$ mHa). Furthermore, Pb ($Z_{\text{eff}}=4$) presents a different scenario in which locality and fixed-node biases are statistically comparable in magnitude ($\epsilon_{\mathrm{loc}} = 1.18(14)$ mHa vs $\epsilon_{\mathrm{FN}} = 1.15(15)$ mHa). This inversion indicate that the non-local ECP angular momentum projectors interact with the diffuse $6p$ valence shell, making the localization approximation the primary error source. While the absolute magnitude of $\epsilon_{\mathrm{loc}}$ depends on the core radius and spatial smoothness of the specific ECP formulation~\cite{Krogel2017,Dzubak2017}, this qualitative error hierarchy is expected to reflect fundamental topological constraints of the single-reference valence nodal surface, suggesting it likely holds broadly across semi-local ECP constructions.

The combined DMC bias ($\eta$) for Tl exhibits a strong dependence on core-valence partitioning, yielding $\eta = 0.98(49)\%$ for $Z_{\text{eff}}=3$, $4.32(21)\%$ for $Z_{\text{eff}}=13$, and $3.74(35)\%$ for $Z_{\text{eff}}=21$. Error partitioning reveals that the elevated bias in the intermediate $Z_{\text{eff}}=13$ space stems primarily from the locality error ($\eta_{\text{loc}} = 1.60(18)\%$), which is three times larger than in the $Z_{\text{eff}}=21$ space ($\eta_{\text{loc}} = 0.555(76)\%$). This locality escalation arises because the non-local angular momentum projectors of the 68-electron core extend spatially into the region of the diffuse $6p$ valence orbital. In contrast, fixed-node errors for $Z_{\text{eff}}=13$ ($2.72(22)\%$) and $Z_{\text{eff}}=21$ ($3.19(35)\%$) are statistically indistinguishable. However, three-dimensional nodal cross-sections of single-determinant HF wavefunctions for the medium-core ($Z_{\text{eff}}=13$) configuration suggest much more nonlinear and possibly topologically different properties, visible in disconnected nodal surfaces (Fig.~\ref{fig:medium_Tl_node}). This node with apparent cylindrical symmetry, reflects the angular nodal features of the non-pseudized $5d$ semi-core shell and resembles features observed previously in studies of idealized $d-$shells~\cite{bajdich2006,Bajdich2008}. These nodal sub-domains restrict walker diffusion and drive local wavefunction curvature, corroborating the kinetic energy ratio ($\xi$) spike observed in Fig.~\ref{fig:xi_6th_row}. Expanding the active space to $Z_{\text{eff}}=21$ eliminates this topological feature and restores locality errors to the universal small-core floor ($\eta_{\text{loc}} \approx 0.55\%$).

% \clearpage
\section{Conclusions}\label{conclusion}
This work quantifies how pseudopotential design affects many-body wavefunction topology, energy partitioning, and fixed-node DMC limits across sixth-period elements (Cs--Rn) using a single-reference Slater-Jastrow trial wavefunction. By combining high-level coupled-cluster and configuration interaction CBS limits with QMC, we characterize the departure of the pseudopotential kinetic-to-total energy ratio ($\xi$) from the all-electron non-relativistic limit ($\xi = 100\%$). Across the series, $\xi$ remains confined between $25.8\%$ and $37.6\%$, governed by the core-size formulation and periodic orbital extent. For a given element, expanding the correlated valence space to include semi-core shells (transitioning from medium- to small-core ECPs) systematically drives $\xi$ downward. This indicates that exposing deep semi-core potential wells increases the magnitude of the total energy faster than the corresponding kinetic energy escalates. Across the period, $\xi$ exhibits distinct block-dependent behavior: the $5d$ block (Hf–Hg) forms a broad plateau near $34\%$ that peaks at hafnium due to the lanthanide contraction. Conversely, $\xi$ declines sharply across the $6p$ block (Tl–Rn) as the diffuse spatial extent of the valence shell dominates the energy partitioning.

The combined DMC bias $\eta$ remains relatively constant across the $5d$ transition metals, but escalates rapidly in the $6p$ block toward the closed-shell noble gas limit, peaking at $5.42(16)\%$ in radon where a single-reference Slater determinant struggles to capture high-density correlation effects. Across most sixth-period elements, fixed-node errors ($\epsilon_{\text{FN}}$) exceed locality errors ($\epsilon_{\text{loc}}$), confirming that trial wavefunction nodal topology remains the primary constraint. However, a locality error inversion occurs at Bi ($\epsilon_{\text{loc}} = 2.09(25)\text{ mHa}$ vs. $\epsilon_{\text{FN}} = 1.20(17)\text{ mHa}$), while Pb exhibits comparable error magnitudes. This inversion indicates that non-local ECP angular-momentum projectors interact strongly with the diffuse $6p$ valence shell, making the localization approximation (T-moves) the dominant error source in these heavy main-group systems.

Beyond macro-scale periodic trends, the core-valence boundary ($Z_{\text{eff}}$) directly alters trial wavefunction nodal topology and locality error. In thallium, the large-core ($Z_{\text{eff}}=3$) and small-core ($Z_{\text{eff}}=21$) configurations yield continuously connected boundaries in 3D nodal cross-sections. In contrast, the medium-core ($Z_{\text{eff}}=13$) configuration forms an isolated toroidal nodal domain near the nucleus, confining walker diffusion and increasing local wavefunction curvature. Error partitioning reveals that the DMC bias in $Z_{\text{eff}}=13$ thallium is driven by locality error ($\eta_{\text{loc}} = 1.60(18)\%$) due to non-local projector overlap with diffuse valence density, whereas fixed-node errors for $Z_{\text{eff}}=13$ ($2.72(22)\%$) and $Z_{\text{eff}}=21$ ($3.19(35)\%$) remain statistically indistinguishable.

In summary, this systematic assessment completes a unified, periodic-table-wide many-body database of ccECP references spanning hydrogen to radon. Delivering these highly resolved benchmarks establishes a physical framework to quantifying the performance of single-determinant Slater–Jastrow trial wavefunctions under the fixed-node approximation across the Cs–Rn series and provides an essential baseline for future developments in systematically improved many-body wavefunction ansatzes and high-precision quantum Monte Carlo applications.
%%%~~~~~~~~~~~~~~~~~~~~~~~~~~~~~~~~~~~~~~~~~~~~~~~~~~~
%%%~~~~~~~~~~~~~~~~~~~~~~~~~~~~~~~~~~~~~~~~~~~~~~~~~~~
% \clearpage
% \input{supporting.tex}
%%%~~~~~~~~~~~~~~~~~~~~~~~~~~~~~~~~~~~~~~~~~~~~~~~~~~~
% \clearpage
% \newpage
\section*{Acknowledgments}
The authors thank Paul R. C. Kent for reading the manuscript and providing helpful suggestions.

This work has been supported by the U.S. Department of Energy, Office of Science, Basic Energy Sciences, Materials Sciences and Engineering Division, as part of the Computational Materials Sciences Program and Center for Predictive Simulation of Functional Materials.

This research used resources of the National Energy Research Scientific Computing Center (NERSC), a U.S. Department of Energy Office of Science User Facility operated under Contract No. DE-AC02-05CH11231.

An award of computer time was provided by the Innovative and Novel Computational Impact on Theory and Experiment (INCITE) program.

This research used resources of the Oak Ridge Leadership Computing Facility, which is a DOE Office of Science User Facility supported under Contract No. DE-AC05-00OR22725.

This paper describes objective technical results and analysis. Any subjective views or opinions that might be expressed in the paper do not necessarily represent the views of the U.S. Department of Energy or the United States Government. 

%%%~~~~~~~~~~~~~~~~~~~~~~~~~~~~~~~~~~~~~~~~~~~~~~~~~~~
\section*{Conflict of Interest}
The authors have no conflicts to disclose.
%%%~~~~~~~~~~~~~~~~~~~~~~~~~~~~~~~~~~~~~~~~~~~~~~~~~~~
\section*{Data Availability} 
The data supporting this study are available within the article and its Supplementary Material. The complete ccECP library, comprising semi-local parameters, and basis sets in various formats is accessible at Ref.\cite{pseudopotentiallibrary}. Raw input/output archives are hosted by the Material Data Facility \cite{Blaiszik2016, Blaiszik2019} at Ref. \cite{mdf-Cs-Rn_data}.

%%%~~~~~~~~~~~~~~~~~~~~~~~~~~~~~~~~~~~~~~~~~~~~~~~~~~~
\section*{REFERENCES}
\bibliographystyle{apsrev4-2}
\bibliography{main.bib}

@article{Ihm1979,
doi = {10.1088/0022-3719/12/21/009},
url = {https://doi.org/10.1088/0022-3719/12/21/009},
year = {1979},
month = {nov},
publisher = {},
volume = {12},
number = {21},
pages = {4409},
author = {J Ihm and A Zunger and M L Cohen},
title = {Momentum-space formalism for the total energy of solids},
journal = {Journal of Physics C: Solid State Physics},
}

@Article{Bitter2010,
author={Bitter, T.
and Wang, S. G.
and Ruedenberg, K.
and Schwarz, W. H. E.},
title={Toward a physical understanding of electron-sharing two-center bonds. II. Pseudo-potential based analysis of diatomic molecules},
journal={Theoretical Chemistry Accounts},
year={2010},
month={Oct},
day={01},
volume={127},
number={3},
pages={237-257},
issn={1432-2234},
doi={10.1007/s00214-010-0758-y},
url={https://doi.org/10.1007/s00214-010-0758-y}
}

@article{Pickett1989,
title = {Pseudopotential methods in condensed matter applications},
journal = {Computer Physics Reports},
volume = {9},
number = {3},
pages = {115-197},
year = {1989},
issn = {0167-7977},
doi = {https://doi.org/10.1016/0167-7977(89)90002-6},
url = {https://www.sciencedirect.com/science/article/pii/0167797789900026},
author = {Warren E. Pickett},
}

@article{Schwerdtfeger2011,
    author = {Schwerdtfeger, Peter and Assadollahzadeh, Behnam and Rohrmann, Urban and Schäfer, Rolf and Cheeseman, James R.},
    title = {Breakdown of the pseudopotential approximation for magnetizabilities and electric multipole moments: Test calculations for Au, AuF, and Snn cluster (n ⩽ 20)},
    journal = {The Journal of Chemical Physics},
    volume = {134},
    number = {20},
    pages = {204102},
    year = {2011},
    month = {05},
    issn = {0021-9606},
    doi = {10.1063/1.3591338},
    url = {https://doi.org/10.1063/1.3591338},
}

@Article{Kevin2012,
  title = {Impact of electron density on the fixed-node errors in Quantum Monte Carlo of atomic systems},
  author = {K.M. Rasch and L. Mitas},
  journal = {Chem. Phys. Lett.},
  year = {2012},
  pages = {59 - 62},
  volume = {528},
  doi = {https://doi.org/10.1016/j.cplett.2012.01.016},
  issn = {0009-2614},
  url = {http://www.sciencedirect.com/science/article/pii/S0009261412000462}
}

@article{kevin2014,
    author = {Rasch, Kevin M. and Hu, Shuming and Mitas, Lubos},
    title = {Communication: Fixed-node errors in quantum Monte Carlo: Interplay of electron density and node nonlinearities},
    journal = {The Journal of Chemical Physics},
    volume = {140},
    number = {4},
    pages = {041102},
    year = {2014},
    month = {01},
    issn = {0021-9606},
    doi = {10.1063/1.4862496},
    url = {https://doi.org/10.1063/1.4862496},
}

@article{bajdich2006,
  title = {Pfaffian Pairing Wave Functions in Electronic-Structure Quantum Monte Carlo Simulations},
  author = {Bajdich, M. and Mitas, L. and Drobn\'y, G. and Wagner, L. K. and Schmidt, K. E.},
  journal = {Phys. Rev. Lett.},
  volume = {96},
  issue = {13},
  pages = {130201},
  numpages = {4},
  year = {2006},
  month = {Apr},
  publisher = {American Physical Society},
  doi = {10.1103/PhysRevLett.96.130201},
  url = {https://link.aps.org/doi/10.1103/PhysRevLett.96.130201}
}

@article{Bajdich2008,
  title = {Pfaffian pairing and backflow wavefunctions for electronic structure quantum Monte Carlo methods},
  author = {Bajdich, M. and Mitas, L. and Wagner, L. K. and Schmidt, K. E.},
  journal = {Phys. Rev. B},
  volume = {77},
  issue = {11},
  pages = {115112},
  numpages = {13},
  year = {2008},
  month = {Mar},
  publisher = {American Physical Society},
  doi = {10.1103/PhysRevB.77.115112},
  url = {https://link.aps.org/doi/10.1103/PhysRevB.77.115112}
}

@article{Hurley1987,
    author = {Hurley, M. M. and Christiansen, P. A.},
    title = {Relativistic effective potentials in quantum Monte Carlo calculations},
    journal = {The Journal of Chemical Physics},
    volume = {86},
    number = {2},
    pages = {1069-1070},
    year = {1987},
    month = {01},
    issn = {0021-9606},
    doi = {10.1063/1.452294},
    url = {https://doi.org/10.1063/1.452294},
}

@article{Mitas1991,
    author = {Mitáš, Luboš and Shirley, Eric L. and Ceperley, David M.},
    title = {Nonlocal pseudopotentials and diffusion Monte Carlo},
    journal = {The Journal of Chemical Physics},
    volume = {95},
    number = {5},
    pages = {3467-3475},
    year = {1991},
    month = {09},
    issn = {0021-9606},
    doi = {10.1063/1.460849},
    url = {https://doi.org/10.1063/1.460849},
}

@article{Krogel2017,
    author = {Krogel, Jaron T. and Kent, P. R. C.},
    title = {Magnitude of pseudopotential localization errors in fixed node diffusion quantum Monte Carlo},
    journal = {The Journal of Chemical Physics},
    volume = {146},
    number = {24},
    pages = {244101},
    year = {2017},
    month = {06},
    issn = {0021-9606},
    doi = {10.1063/1.4986951},
    url = {https://doi.org/10.1063/1.4986951},
}

@article{Dzubak2017,
    author = {Dzubak, Allison L. and Krogel, Jaron T. and Reboredo, Fernando A.},
    title = {Quantitative estimation of localization errors of 3d transition metal pseudopotentials in diffusion Monte Carlo},
    journal = {The Journal of Chemical Physics},
    volume = {147},
    number = {2},
    pages = {024102},
    year = {2017},
    month = {07},
    issn = {0021-9606},
    doi = {10.1063/1.4991414},
    url = {https://doi.org/10.1063/1.4991414},
}

@article{casula2005,
  title = {Diffusion Monte Carlo Method with Lattice Regularization},
  author = {Casula, Michele and Filippi, Claudia and Sorella, Sandro},
  journal = {Phys. Rev. Lett.},
  volume = {95},
  issue = {10},
  pages = {100201},
  numpages = {4},
  year = {2005},
  month = {Sep},
  publisher = {American Physical Society},
  doi = {10.1103/PhysRevLett.95.100201},
  url = {https://link.aps.org/doi/10.1103/PhysRevLett.95.100201}
}

@article{Haihan2022,
title = {A quantum Monte Carlo study of systems with effective core potentials and node nonlinearities},
journal = {Chemical Physics},
volume = {554},
pages = {111402},
year = {2022},
issn = {0301-0104},
doi = {https://doi.org/10.1016/j.chemphys.2021.111402},
url = {https://www.sciencedirect.com/science/article/pii/S030101042100313X},
author = {Haihan Zhou and Anthony Scemama and Guangming Wang and Abdulgani Annaberdiyev and Benjamin Kincaid and Michel Caffarel and Lubos Mitas},
}

@article{jaehyung2015,
    author = {Yu, Jaehyung and Wagner, Lucas K. and Ertekin, Elif},
    title = {Towards a systematic assessment of errors in diffusion Monte Carlo calculations of semiconductors: Case study of zinc selenide and zinc oxide},
    journal = {The Journal of Chemical Physics},
    volume = {143},
    number = {22},
    pages = {224707},
    year = {2015},
    month = {12},
    issn = {0021-9606},
    doi = {10.1063/1.4937421},
    url = {https://doi.org/10.1063/1.4937421},
}

@article{casula2006,
  title = {Beyond the locality approximation in the standard diffusion Monte Carlo method},
  author = {Casula, Michele},
  journal = {Phys. Rev. B},
  volume = {74},
  issue = {16},
  pages = {161102(R)},
  numpages = {4},
  year = {2006},
  month = {Oct},
  publisher = {American Physical Society},
  doi = {10.1103/PhysRevB.74.161102},
  url = {https://link.aps.org/doi/10.1103/PhysRevB.74.161102}
}

@article{casula2010,
    author = {Casula, Michele and Moroni, Saverio and Sorella, Sandro and Filippi, Claudia},
    title = {Size-consistent variational approaches to nonlocal pseudopotentials: Standard and lattice regularized diffusion Monte Carlo methods revisited},
    journal = {The Journal of Chemical Physics},
    volume = {132},
    number = {15},
    pages = {154113},
    year = {2010},
    month = {04},
    issn = {0021-9606},
    doi = {10.1063/1.3380831},
    url = {https://doi.org/10.1063/1.3380831},
}

@Article{varandas2007,
  title = {Accurate ab initio-based molecular potentials: from extrapolation methods to global modelling},
  author = {A J C Varandas},
  journal = {Phys. Scr.},
  year = {2007},
  month = {aug},
  number = {3},
  pages = {C28--C35},
  volume = {76},
  doi = {10.1088/0031-8949/76/3/n04},
  publisher = {{IOP} Publishing},
  url = {https://doi.org/10.1088%2F0031-8949%2F76%2F3%2Fn04}
}

@Article{Bennett2017,
  author  = {Bennett, M. Chandler and Melton, Cody A. and Annaberdiyev, Abdulgani and Wang, Guangming and Shulenburger, Luke and Mitas, Lubos},
  journal = {J. Chem. Phys.},
  title   = {A New Generation of Effective Core Potentials for Correlated Calculations},
  year    = {2017},
  issn    = {0021-9606},
  month   = dec,
  number  = {22},
  pages   = {224106},
  volume  = {147},
  doi     = {10.1063/1.4995643},
  urldate = {2019-12-17},
}

@Article{Bennett2018,
  author     = {Bennett, M. Chandler and Wang, Guangming and Annaberdiyev, Abdulgani and Melton, Cody A. and Shulenburger, Luke and Mitas, Lubos},
  journal    = {J. Chem. Phys.},
  title      = {A New Generation of Effective Core Potentials from Correlated Calculations: 2nd Row Elements},
  year       = {2018},
  issn       = {0021-9606},
  month      = sep,
  number     = {10},
  pages      = {104108},
  volume     = {149},
  doi        = {10.1063/1.5038135},
  shorttitle = {A New Generation of Effective Core Potentials from Correlated Calculations},
  urldate    = {2019-12-17},
}

@Article{Annaberdiyev2018,
  author     = {Annaberdiyev, Abdulgani and Wang, Guangming and Melton, Cody A. and Bennett, M. Chandler and Shulenburger, Luke and Mitas, Lubos},
  journal    = {J. Chem. Phys.},
  title      = {A New Generation of Effective Core Potentials from Correlated Calculations: 3d Transition Metal Series},
  year       = {2018},
  issn       = {0021-9606},
  month      = oct,
  number     = {13},
  pages      = {134108},
  volume     = {149},
  doi        = {10.1063/1.5040472},
  shorttitle = {A New Generation of Effective Core Potentials from Correlated Calculations},
  urldate    = {2019-12-17},
}

@Article{Wang2019,
  author     = {Wang, Guangming and Annaberdiyev, Abdulgani and Melton, Cody A. and Bennett, M. Chandler and Shulenburger, Luke and Mitas, Lubos},
  journal    = {J. Chem. Phys.},
  title      = {A New Generation of Effective Core Potentials from Correlated Calculations: 4s and 4p Main Group Elements and First Row Additions},
  year       = {2019},
  issn       = {0021-9606},
  month      = oct,
  number     = {14},
  pages      = {144110},
  volume     = {151},
  doi        = {10.1063/1.5121006},
  publisher  = {{American Institute of Physics}},
  shorttitle = {A New Generation of Effective Core Potentials from Correlated Calculations},
  urldate    = {2021-01-18},
}

@Article{Wang2022,
  author     = {Wang, Guangming and Kincaid, Benjamin and Zhou, Haihan and Annaberdiyev, Abdulgani and Bennett, M. Chandler and Krogel, Jaron T. and Mitas, Lubos},
  journal    = {J. Chem. Phys.},
  title      = {A New Generation of Effective Core Potentials from Correlated and Spin\textendash Orbit Calculations: {{Selected}} Heavy Elements},
  year       = {2022},
  issn       = {0021-9606},
  month      = aug,
  number     = {5},
  pages      = {054101},
  volume     = {157},
  doi        = {10.1063/5.0087300},
  publisher  = {{American Institute of Physics}},
  shorttitle = {A New Generation of Effective Core Potentials from Correlated and Spin\textendash Orbit Calculations},
  urldate    = {2022-08-02},
}

@Article{Kincaid2022,
  author    = {Kincaid, Benjamin and Wang, Guangming and Zhou, Haihan and Mitas, Lubos},
  journal   = {The Journal of Chemical Physics},
  title     = {Correlation Consistent Effective Core Potentials for Late 3d Transition Metals Adapted for Plane Wave Calculations},
  year      = {2022},
  issn      = {0021-9606},
  month     = nov,
  number    = {17},
  pages     = {174307},
  volume    = {157},
  doi       = {10.1063/5.0109098},
  publisher = {{American Institute of Physics}},
  urldate   = {2022-11-14},
}

@article{annaberdiyev2023,
  title = {The Role of Electron Correlations in the Electronic Structure of Putative {{Chern}} Magnet {{TbMn6Sn6}}},
  author = {Annaberdiyev, Abdulgani and Mandal, Subhasish and Mitas, Lubos and Krogel, Jaron T. and Ganesh, Panchapakesan},
  year = {2023},
  month = sep,
  journal = {npj Quantum Materials},
  volume = {8},
  number = {1},
  pages = {1--13},
  publisher = {{Nature Publishing Group}},
  issn = {2397-4648},
  doi = {10.1038/s41535-023-00583-6},
  urldate = {2023-09-27},
  copyright = {2023 Springer Nature Limited},
  langid = {english}
}

@article{Haihan2024,
    author = {Zhou, Haihan and Kincaid, Benjamin and Wang, Guangming and Annaberdiyev, Abdulgani and Ganesh, Panchapakesan and Mitas, Lubos},
    title = {A new generation of effective core potentials: Selected lanthanides and heavy elements},
    journal = {The Journal of Chemical Physics},
    volume = {160},
    number = {8},
    pages = {084302},
    year = {2024},
    month = {02},
    issn = {0021-9606},
    doi = {10.1063/5.0180057},
    url = {https://doi.org/10.1063/5.0180057},

}

@article{Omar2025,
    author = {Madany, Omar and Kincaid, Benjamin and Shaikh, Aqsa and Morningstar, Elizabeth and Mitas, Lubos},
    title = {A new generation of effective core potentials: Selected lanthanides and heavy elements II},
    journal = {The Journal of Chemical Physics},
    volume = {163},
    number = {11},
    pages = {114108},
    year = {2025},
    month = {09},
    issn = {0021-9606},
    doi = {10.1063/5.0285320},
    url = {https://doi.org/10.1063/5.0285320},
}

@misc{Omar2026,
      title={A new generation of effective core potentials: Selected heavy 5d and 6p elements}, 
      author={Omar Madany and Lubos Mitas},
      year={2026},
      eprint={2609.10853},
      archivePrefix={arXiv},
      primaryClass={physics.comp-ph},
      url={https://arxiv.org/abs/2609.10853}, 
}

@Misc{pseudopotentiallibrary,
  howpublished = {\url{https://pseudopotentiallibrary.org}},
  note         = {Accessed: 2023-08-02},
  title        = {Pseudopotential {Library}: {A} community website for pseudopotentials/effective core potentials developed for high accuracy correlated many-body methods such as quantum {Monte} {Carlo} and quantum chemistry},
}

@article{Pyykko2002,
    author = {Pyykko, Pekka},
    title = {Relativistic effects in structural chemistry},
    journal = {Chemical Reviews},
    volume = {88},
    number = {3},
    pages = {563-594},
    year = {2002},
    month = {05},
    issn = {0009-2665},
    doi = {10.1021/cr00085a006},
    url = {https://doi.org/10.1021/cr00085a006},
}

@article{Dolg2011,
    author = {Dolg, Michael and Cao, Xiaoyan},
    title = {Relativistic Pseudopotentials: Their Development and Scope of Applications},
    journal = {Chemical Reviews},
    volume = {112},
    number = {1},
    pages = {403-480},
    year = {2011},
    month = {09},
    issn = {0009-2665},
    doi = {10.1021/cr2001383},
    url = {https://doi.org/10.1021/cr2001383},
}

@article{acc_engI,
  author    = {Annaberdiyev, Abdulgani and Melton, Cody A. and Bennett, M. Chandler and Wang, Guangming and Mitas, Lubos},
  journal   = {J. Chem. Theory Comput.},
  title     = {Accurate {{Atomic Correlation}} and {{Total Energies}} for {{Correlation Consistent Effective Core Potentials}}},
  year      = {2020},
  issn      = {1549-9618},
  month     = mar,
  number    = {3},
  pages     = {1482--1502},
  volume    = {16},
  doi       = {10.1021/acs.jctc.9b00962},
  publisher = {{American Chemical Society}},
  urldate   = {2020-03-11},
}

@article{acc_engII,
author = {Shaikh, Aqsa and Madany, Omar and Kincaid, Benjamin and Mitas, Lubos},
title = {Accurate Atomic Correlation and Total Energies for Correlation-Consistent Effective Core Potentials II: Rb–Xe Elements},
journal = {Journal of Chemical Theory and Computation},
volume = {22},
number = {9},
pages = {4716-4727},
year = {2026},
doi = {10.1021/acs.jctc.5c02129},
URL = {https://doi.org/10.1021/acs.jctc.5c02129},
}

@article{Desclaux1973,
title = {Relativistic Dirac-Fock expectation values for atoms with Z = 1 to Z = 120},
journal = {Atomic Data and Nuclear Data Tables},
volume = {12},
number = {4},
pages = {311-406},
year = {1973},
issn = {0092-640X},
doi = {https://doi.org/10.1016/0092-640X(73)90020-X},
url = {https://www.sciencedirect.com/science/article/pii/0092640X7390020X},
author = {J.P. Desclaux},
}

@article{Kotochigova1997,
  title = {Local-density-functional calculations of the energy of atoms},
  author = {Kotochigova, Svetlana and Levine, Zachary H. and Shirley, Eric L. and Stiles, M. D. and Clark, Charles W.},
  journal = {Phys. Rev. A},
  volume = {55},
  issue = {1},
  pages = {191--199},
  numpages = {0},
  year = {1997},
  month = {Jan},
  publisher = {American Physical Society},
  doi = {10.1103/PhysRevA.55.191},
  url = {https://link.aps.org/doi/10.1103/PhysRevA.55.191}
}

@article{Rolfes2020,
author = {Rolfes, Julian D. and Neese, Frank and Pantazis, Dimitrios A.},
title = {All-electron scalar relativistic basis sets for the elements Rb–Xe},
journal = {Journal of Computational Chemistry},
volume = {41},
number = {20},
pages = {1842-1849},
doi = {https://doi.org/10.1002/jcc.26355},
url = {https://onlinelibrary.wiley.com/doi/abs/10.1002/jcc.26355},
}

@article{Ishikawa1994,
  title = {Relativistic many-body perturbation theory based on the no-pair Dirac-Coulomb-Breit Hamiltonian: Relativistic correlation energies for the noble-gas sequence through Rn (Z=86), the group-IIB atoms through Hg, and the ions of Ne isoelectronic sequence},
  author = {Ishikawa, Yasuyuki and Koc, Konrad},
  journal = {Phys. Rev. A},
  volume = {50},
  issue = {6},
  pages = {4733--4742},
  numpages = {0},
  year = {1994},
  month = {Dec},
  publisher = {American Physical Society},
  doi = {10.1103/PhysRevA.50.4733},
  url = {https://link.aps.org/doi/10.1103/PhysRevA.50.4733}
}

@article{Kumar2021,
  title = {Relativistic coupled-cluster calculation of the electric dipole polarizability and correlation energy of Cn, ${\mathrm{Nh}}^{+}$, and Og: Correlation effects from lighter to superheavy elements},
  author = {Kumar, Ravi and Chattopadhyay, S. and Angom, D. and Mani, B. K.},
  journal = {Phys. Rev. A},
  volume = {103},
  issue = {6},
  pages = {062803},
  numpages = {16},
  year = {2021},
  month = {Jun},
  publisher = {American Physical Society},
  doi = {10.1103/PhysRevA.103.062803},
  url = {https://link.aps.org/doi/10.1103/PhysRevA.103.062803}
}

@article{Hoyer2023,
    author = {Hoyer, Chad E. and Lu, Lixin and Hu, Hang and Shumilov, Kirill D. and Sun, Shichao and Knecht, Stefan and Li, Xiaosong},
    title = {Correlated Dirac–Coulomb–Breit multiconfigurational self-consistent-field methods},
    journal = {The Journal of Chemical Physics},
    volume = {158},
    number = {4},
    pages = {044101},
    year = {2023},
    month = {01},
    issn = {0021-9606},
    doi = {10.1063/5.0133741},
    url = {https://doi.org/10.1063/5.0133741},
}

@article{Dzuba2024,
  title = {Calculation of the correlation, relativistic, and QED corrections to the total electron binding energy in atoms and their nuclear charge dependence},
  author = {Dzuba, V. A. and Flambaum, V. V. and Afanasjev, A. V.},
  journal = {Phys. Rev. A},
  volume = {110},
  issue = {5},
  pages = {052810},
  numpages = {10},
  year = {2024},
  month = {Nov},
  publisher = {American Physical Society},
  doi = {10.1103/PhysRevA.110.052810},
  url = {https://link.aps.org/doi/10.1103/PhysRevA.110.052810}
}

@article{Chakravorty1993,
  title = {Ground-state correlation energies for atomic ions with 3 to 18 electrons},
  author = {Chakravorty, Subhas J. and Gwaltney, Steven R. and Davidson, Ernest R. and Parpia, Farid A. and p Fischer, Charlotte Froese},
  journal = {Phys. Rev. A},
  volume = {47},
  issue = {5},
  pages = {3649--3670},
  numpages = {0},
  year = {1993},
  month = {May},
  publisher = {American Physical Society},
  doi = {10.1103/PhysRevA.47.3649},
  url = {https://link.aps.org/doi/10.1103/PhysRevA.47.3649}
}

@article{Luchow1996,
    author = {Lüchow, Arne and Anderson, James B.},
    title = {First‐row hydrides: Dissociation and ground state energies using quantum Monte Carlo},
    journal = {The Journal of Chemical Physics},
    volume = {105},
    number = {17},
    pages = {7573-7578},
    year = {1996},
    month = {11},
    issn = {0021-9606},
    doi = {10.1063/1.472584},
    url = {https://doi.org/10.1063/1.472584},
}

@article{Jitrik1997,
  title = {Atomic configuration interaction and studies of He, Li, Be, and Ne ground states},
  author = {Jitrik, Oliverio and Bunge, Carlos F.},
  journal = {Phys. Rev. A},
  volume = {56},
  issue = {4},
  pages = {2614--2623},
  numpages = {0},
  year = {1997},
  month = {Oct},
  publisher = {American Physical Society},
  doi = {10.1103/PhysRevA.56.2614},
  url = {https://link.aps.org/doi/10.1103/PhysRevA.56.2614}
}

@article{Booth2010,
    author = {Booth, George H. and Alavi, Ali},
    title = {Approaching chemical accuracy using full configuration-interaction quantum Monte Carlo: A study of ionization potentials},
    journal = {The Journal of Chemical Physics},
    volume = {132},
    number = {17},
    pages = {174104},
    year = {2010},
    month = {05},
    issn = {0021-9606},
    doi = {10.1063/1.3407895},
    url = {https://doi.org/10.1063/1.3407895},
}

@article{Cleland2011,
    author = {Cleland, D. M. and Booth, George H. and Alavi, Ali},
    title = {A study of electron affinities using the initiator approach to full configuration interaction quantum Monte Carlo},
    journal = {The Journal of Chemical Physics},
    volume = {134},
    number = {2},
    pages = {024112},
    year = {2011},
    month = {01},
    issn = {0021-9606},
    doi = {10.1063/1.3525712},
    url = {https://doi.org/10.1063/1.3525712},
}

@article{McCarthy2011,
    author = {McCarthy, Shane P. and Thakkar, Ajit J.},
    title = {Accurate all-electron correlation energies for the closed-shell atoms from Ar to Rn and their relationship to the corresponding MP2 correlation energies},
    journal = {The Journal of Chemical Physics},
    volume = {134},
    number = {4},
    pages = {044102},
    year = {2011},
    month = {01},
    issn = {0021-9606},
    doi = {10.1063/1.3547262},
    url = {https://doi.org/10.1063/1.3547262},
}

@article{McCarthy2012,
    author = {McCarthy, Shane P. and Thakkar, Ajit J.},
    title = {When does the non-variational nature of second-order Møller-Plesset energies manifest itself? All-electron correlation energies for open-shell atoms from K to Br},
    journal = {The Journal of Chemical Physics},
    volume = {136},
    number = {5},
    pages = {054107},
    year = {2012},
    month = {02},
    issn = {0021-9606},
    doi = {10.1063/1.3679969},
    url = {https://doi.org/10.1063/1.3679969},
}

@article{Buendia2013,
title = {Quantum Monte Carlo ionization potential and electron affinity for transition metal atoms},
journal = {Chemical Physics Letters},
volume = {559},
pages = {12-17},
year = {2013},
issn = {0009-2614},
doi = {https://doi.org/10.1016/j.cplett.2012.12.055},
url = {https://www.sciencedirect.com/science/article/pii/S0009261413000079},
author = {E. Buendía and F.J. Gálvez and P. Maldonado and A. Sarsa},
}

@article{Ranasinghe2013,
    author = {Ranasinghe, Duminda S. and Petersson, George A.},
    title = {CCSD(T)/CBS atomic and molecular benchmarks for H through Ar},
    journal = {The Journal of Chemical Physics},
    volume = {138},
    number = {14},
    pages = {144104},
    year = {2013},
    month = {04},
    issn = {0021-9606},
    doi = {10.1063/1.4798707},
    url = {https://doi.org/10.1063/1.4798707},
}

@article{Scemama2014,
    author = {Scemama, A. and Applencourt, T. and Giner, E. and Caffarel, M.},
    title = {Accurate nonrelativistic ground-state energies of 3d transition metal atoms},
    journal = {The Journal of Chemical Physics},
    volume = {141},
    number = {24},
    pages = {244110},
    year = {2014},
    month = {12},
    issn = {0021-9606},
    doi = {10.1063/1.4903985},
    url = {https://doi.org/10.1063/1.4903985},
}

@article{Ranasinghe2015,
    author = {Ranasinghe, Duminda S. and Frisch, Michael J. and Petersson, George A.},
    title = {Core-core and core-valence correlation energy atomic and molecular benchmarks for Li through Ar},
    journal = {The Journal of Chemical Physics},
    volume = {143},
    number = {21},
    pages = {214110},
    year = {2015},
    month = {12},
    issn = {0021-9606},
    doi = {10.1063/1.4935972},
    url = {https://doi.org/10.1063/1.4935972},
}

@article{Li2019,
	title = {Comparing many-body approaches against the helium atom exact solution},
	pages = {040},
	author = {Li, Jing and Drummond, N. D. and Schuck, Peter and Olevano, Valerio},
	journal = {SciPost Phys.},
	volume = {6},
	year = {2019},
	publisher = {SciPost},
	doi = {10.21468/SciPostPhys.6.4.040},
	url = {https://scipost.org/10.21468/SciPostPhys.6.4.040}
}

@article{Peng2023,
    author = {Peng, Hao and Yang, Sixian and Jiang, Hong and Weng, Hongming and Ren, Xinguo},
    title = {Basis-Set-Error-Free
Random-Phase Approximation Correlation
Energies for Atoms Based on the Sternheimer Equation},
    journal = {Journal of Chemical Theory and Computation},
    volume = {19},
    number = {20},
    pages = {7199-7214},
    year = {2023},
    month = {10},
    issn = {1549-9618},
    doi = {10.1021/acs.jctc.3c00668},
    url = {https://doi.org/10.1021/acs.jctc.3c00668},
}

@article{molpro-2012,
author = {Werner, Hans-Joachim and Knowles, Peter J. and Knizia, Gerald and Manby, Frederick R. and Schütz, Martin},
title = {Molpro: a general-purpose quantum chemistry program package},
journal = {WIREs Computational Molecular Science},
volume = {2},
number = {2},
pages = {242-253},
doi = {https://doi.org/10.1002/wcms.82},
url = {https://wires.onlinelibrary.wiley.com/doi/abs/10.1002/wcms.82},
}

@article{molpro-2020,
    author = {Werner, Hans-Joachim and Knowles, Peter J. and Manby, Frederick R. and Black, Joshua A. and Doll, Klaus and Heßelmann, Andreas and Kats, Daniel and Köhn, Andreas and Korona, Tatiana and Kreplin, David A. and Ma, Qianli and Miller, Thomas F., III and Mitrushchenkov, Alexander and Peterson, Kirk A. and Polyak, Iakov and Rauhut, Guntram and Sibaev, Marat},
    title = {The Molpro quantum chemistry package},
    journal = {The Journal of Chemical Physics},
    volume = {152},
    number = {14},
    pages = {144107},
    year = {2020},
    month = {04},
    issn = {0021-9606},
    doi = {10.1063/5.0005081},
    url = {https://doi.org/10.1063/5.0005081},
}

@misc{molpro-3,
title={MOLPRO, version , a package of ab initio programs},
author={

H.-J. Werner and
P. J. Knowles and

P. Celani and
W. Gy\"orffy and
A. Hesselmann and
D. Kats and
G. Knizia and
A. K\"ohn and
T. Korona and
D. Kreplin and
R. Lindh and
Q. Ma and
F. R. Manby and
A. Mitrushenkov and
G. Rauhut and
M. {Sch\"{u}tz} and
K. R. Shamasundar and

T. B. Adler and
R. D. Amos and
S. J. Bennie and
A. Bernhardsson and
A. Berning and
J. A. Black and
P. J. Bygrave and
R. Cimiraglia and
D. L. Cooper and
D. Coughtrie and
M. J. O. Deegan and
A. J. Dobbyn and
K. Doll and 
M. Dornbach and
F. Eckert and
S. Erfort and
E. Goll and
C. Hampel and
G. Hetzer and
J. G. Hill and
M. Hodges and 
T. Hrenar and
G. Jansen and
C. K\"oppl and
C. Kollmar and
S. J. R. Lee and
Y. Liu and
A. W. Lloyd and
R. A. Mata and
A. J. May and
B. Mussard and
S. J. McNicholas and
W. Meyer and
T. F. {Miller III} and
M. E. Mura and
A. Nicklass and
D. P. O'Neill and
P. Palmieri and
D. Peng and
K. A. Peterson and
K. Pfl\"uger and
R. Pitzer and
I. Polyak and
M. Reiher and
J. O. Richardson and
J. B. Robinson and
B. Schr\"oder and
M. Schwilk and 
T. Shiozaki and
M. Sibaev and
H. Stoll and
A. J. Stone and
R. Tarroni and
T. Thorsteinsson and
J. Toulouse and
M. Wang and
M. Welborn and 
B. Ziegler

},
note={see https://www.molpro.net/},
}

@article{molpro-cc,
    author = {Knowles, Peter J. and Hampel, Claudia and Werner, Hans-Joachim},
    title = {Erratum: “Coupled cluster theory for high spin, open shell reference wave functions” [ J. Chem. Phys. 99, 5219 (1993)]},
    journal = {The Journal of Chemical Physics},
    volume = {112},
    number = {6},
    pages = {3106-3107},
    year = {2000},
    month = {02},
    issn = {0021-9606},
    doi = {10.1063/1.480886},
    url = {https://doi.org/10.1063/1.480886},
}

@article{molpro-seward,
    author = {Lindh, R. and Ryu, U. and Liu, B.},
    title = {The reduced multiplication scheme of the Rys quadrature and new recurrence relations for auxiliary function based two‐electron integral evaluation},
    journal = {The Journal of Chemical Physics},
    volume = {95},
    number = {8},
    pages = {5889-5897},
    year = {1991},
    month = {10},
    issn = {0021-9606},
    doi = {10.1063/1.461610},
    url = {https://doi.org/10.1063/1.461610},
}

@article{molpro-mcscf1,
title = {An efficient second-order MC SCF method for long configuration expansions},
journal = {Chemical Physics Letters},
volume = {115},
number = {3},
pages = {259-267},
year = {1985},
issn = {0009-2614},
doi = {https://doi.org/10.1016/0009-2614(85)80025-7},
url = {https://www.sciencedirect.com/science/article/pii/0009261485800257},
author = {Peter J. Knowles and Hans-Joachim Werner}
}

@article{molpro-mcscf2,
    author = {Kreplin, David A. and Knowles, Peter J. and Werner, Hans-Joachim},
    title = {Second-order MCSCF optimization revisited. I. Improved algorithms for fast and robust second-order CASSCF convergence},
    journal = {The Journal of Chemical Physics},
    volume = {150},
    number = {19},
    pages = {194106},
    year = {2019},
    month = {05},
    issn = {0021-9606},
    doi = {10.1063/1.5094644},
    url = {https://doi.org/10.1063/1.5094644},
}

@article{molpro-fci,
title = {A new determinant-based full configuration interaction method},
journal = {Chemical Physics Letters},
volume = {111},
number = {4},
pages = {315-321},
year = {1984},
issn = {0009-2614},
doi = {https://doi.org/10.1016/0009-2614(84)85513-X},
url = {https://www.sciencedirect.com/science/article/pii/000926148485513X},
author = {P.J. Knowles and N.C. Handy}
}

@article{mrcc-1,
author = {Mester, Dávid and Nagy, P{\'e}ter R. and Csóka, József and Gyevi-Nagy, László and Szabó, P. Bernát and Horváth, R{\'e}ka A. and Petrov, Klára and H{\'e}gely, Bence and Ladóczki, Bence and Samu, Gyula and Lőrincz, Balázs D. and Kállay, Mihály},
title = {Overview of Developments in the MRCC Program System},
journal = {The Journal of Physical Chemistry A},
volume = {129},
number = {8},
pages = {2086-2107},
year = {2025},
doi = {10.1021/acs.jpca.4c07807},
URL = {https://doi.org/10.1021/acs.jpca.4c07807},
}

@Misc{mrcc-2,
  title = {Mrcc, a quantum chemical program suite},
  author = {M. Kallay and P. R. Nagy and Z. Rolik and D. Mester and G. Samu and J. Csontos and J. Csoka and B. P. Szabo and L. Gyevi-Nagy and I. Ladjanszki and L. Szegedy and B. Ladoczki and K. Petrov and M. Farkas and P. D. Mezei and and B. Hegely},
  note = {See also Z. Rolik, L. Szegedy, I. Ladjánszki, B. Ladóczki, and M. Kállay, \textit{J. Chem. Phys.} \textbf{2013}, \textit{139}, 094105},
  url = {www.mrcc.hu},
}

@article{GAMESS,
title = {Recent developments in the general atomic and molecular electronic structure system},
volume = {152},
issn = {0021-9606, 1089-7690},
url = {http://aip.scitation.org/doi/10.1063/5.0005188},
doi = {10.1063/5.0005188},
language = {en},
number = {15},
urldate = {2020-06-18},
journal = {The Journal of Chemical Physics},
author = {Barca, Giuseppe M. J. and Bertoni, Colleen and Carrington, Laura and Datta, Dipayan and De Silva, Nuwan and Deustua, J. Emiliano and Fedorov, Dmitri G. and Gour, Jeffrey R. and Gunina, Anastasia O. and Guidez, Emilie and Harville, Taylor and Irle, Stephan and Ivanic, Joe and Kowalski, Karol and Leang, Sarom S. and Li, Hui and Li, Wei and Lutz, Jesse J. and Magoulas, Ilias and Mato, Joani and Mironov, Vladimir and Nakata, Hiroya and Pham, Buu Q. and Piecuch, Piotr and Poole, David and Pruitt, Spencer R. and Rendell, Alistair P. and Roskop, Luke B. and Ruedenberg, Klaus and Sattasathuchana, Tosaporn and Schmidt, Michael W. and Shen, Jun and Slipchenko, Lyudmila and Sosonkina, Masha and Sundriyal, Vaibhav and Tiwari, Ananta and Galvez Vallejo, Jorge L. and Westheimer, Bryce and Wloch, Marta and Xu, Peng and Zahariev, Federico and Gordon, Mark S.},
month = apr,
year = {2020},
pages = {154102}
}

@article{pyscf1,
author = {Sun, Qiming},
title = {Libcint: An efficient general integral library for Gaussian basis functions},
journal = {Journal of Computational Chemistry},
volume = {36},
number = {22},
pages = {1664-1671},
doi = {https://doi.org/10.1002/jcc.23981},
url = {https://onlinelibrary.wiley.com/doi/abs/10.1002/jcc.23981},
eprint = {https://onlinelibrary.wiley.com/doi/pdf/10.1002/jcc.23981},
year = {2015}
}

@article{pyscf2,
author = {Sun, Qiming and Berkelbach, Timothy C. and Blunt, Nick S. and Booth, George H. and Guo, Sheng and Li, Zhendong and Liu, Junzi and McClain, James D. and Sayfutyarova, Elvira R. and Sharma, Sandeep and Wouters, Sebastian and Chan, Garnet Kin-Lic},
title = {PySCF: the Python-based simulations of chemistry framework},
journal = {WIREs Computational Molecular Science},
volume = {8},
number = {1},
pages = {e1340},
doi = {https://doi.org/10.1002/wcms.1340},
url = {https://wires.onlinelibrary.wiley.com/doi/abs/10.1002/wcms.1340},
eprint = {https://wires.onlinelibrary.wiley.com/doi/pdf/10.1002/wcms.1340},
year = {2018}
}

@article{pyscf3,
    author = {Sun, Qiming and Zhang, Xing and Banerjee, Samragni and Bao, Peng and Barbry, Marc and Blunt, Nick S. and Bogdanov, Nikolay A. and Booth, George H. and Chen, Jia and Cui, Zhi-Hao and Eriksen, Janus J. and Gao, Yang and Guo, Sheng and Hermann, Jan and Hermes, Matthew R. and Koh, Kevin and Koval, Peter and Lehtola, Susi and Li, Zhendong and Liu, Junzi and Mardirossian, Narbe and McClain, James D. and Motta, Mario and Mussard, Bastien and Pham, Hung Q. and Pulkin, Artem and Purwanto, Wirawan and Robinson, Paul J. and Ronca, Enrico and Sayfutyarova, Elvira R. and Scheurer, Maximilian and Schurkus, Henry F. and Smith, James E. T. and Sun, Chong and Sun, Shi-Ning and Upadhyay, Shiv and Wagner, Lucas K. and Wang, Xiao and White, Alec and Whitfield, James Daniel and Williamson, Mark J. and Wouters, Sebastian and Yang, Jun and Yu, Jason M. and Zhu, Tianyu and Berkelbach, Timothy C. and Sharma, Sandeep and Sokolov, Alexander Yu. and Chan, Garnet Kin-Lic},
    title = {Recent developments in the PySCF program package},
    journal = {The Journal of Chemical Physics},
    volume = {153},
    number = {2},
    pages = {024109},
    year = {2020},
    month = {07},
    issn = {0021-9606},
    doi = {10.1063/5.0006074},
    url = {https://doi.org/10.1063/5.0006074},
}

@article{pyscf4,
    author = {Sun, Qiming and Hermes, Matthew R. and Wu, Xiaojie and Zhai, Huanchen and Zhang, Xing and Ahmed, Abdelrahman M. and Aucar, Juan José and Backhouse, Oliver J. and Banerjee, Samragni and Bao, Peng and Bogdanov, Nikolay A. and Bystrom, Kyle and Chapoton, Frédéric and Chen, Ning-Yuan and Chernyshov, Ivan Yu. and Clifford, Helen S. and Cohen-Janes, Sander and Cui, Zhi-Hao and Damour, Yann D. and Dattani, Nike and Dittmer, Linus Bjarne and Ehlert, Sebastian and Eriksen, Janus Juul and Evangelista, Francesco A. and Ewing, Simon A. and Farahvash, Ardavan and Focke, Kevin and Gao, Yang and Gasperich, Kevin E. and Gillispie, Nathan and Greiner, Jonas and Hennefarth, Matthew R. and Hermann, Jan and Hillenbrand, Christopher and Huhtasalo, Joonatan and Ibrahim, Basil and Jangid, Bhavnesh and Javaremi, Alireza Nejati and Jenkins, Andrew J. and Jin, Yu and King, Daniel S. and Kooi, Derk Pieter and Kurian, Jo S. and Larsson, Henrik R. and Lau, Bryan Tak Gwong and Lee, Seunghoon and Lehtola, Susi and Li, Chenghan and Li, Hao and Li, Jiachen and Li, Rui and Li, Shuhang and Lykhin, Aleksandr O. and Mahajan, Ankit and Mauger, Nastasia and Mazo-Sevillano, Pablo del and Moussa, Jonathan and Nakano, Kousuke and Neufeld, Verena A. and Peng, Linqing and Pham, Hung Q. and Pinski, Peter and Pokhilko, Pavel and Pu, Zhichen and Qian, Yubing and Quiton, Stephen Jon and Schulze, Wanja T. and Scott, Thais R. and Seal, Aniruddha and Serna, James D. and Smith, James E. T. and Smyser, Kori E. and Stahl, Terrence and Sun, Chong and Sung, Kevin J. and Trushin, Egor and Upadhyay, Shiv and Vo, Ethan A. and Vogels, Thijs and Wang, Shirong and Wang, Tai and Wang, Xiao and Wang, Xubo and Wang, Yuanheng and Williamson, Mark and Yang, Junjie and Ye, Hong-Zhou and Yeh, Chia-Nan and Yu, Haiyang and Yu, Jincheng and Yu, Victor Wen-zhe and Zhang, Chaoqun and Zhang, Dayou and Zhang, Yichi and Zhao, Zijun and Zhou, Zehao and Zhu, Andrew J. and Zhu, Tianyu and Berkelbach, Timothy C. and Gagliardi, Laura and Sharma, Sandeep and Sokolov, Alexander Yu. and Kin-Lic Chan, Garnet},
    title = {The Python simulations of chemistry framework: 10 years of an open-source quantum chemistry project},
    journal = {The Journal of Chemical Physics},
    volume = {165},
    number = {10},
    pages = {102502},
    year = {2026},
    month = {09},
    issn = {0021-9606},
    doi = {10.1063/5.0337441},
    url = {https://doi.org/10.1063/5.0337441},
}

@Article{Blaiszik2016,
  author     = {Blaiszik, B. and Chard, K. and Pruyne, J. and Ananthakrishnan, R. and Tuecke, S. and Foster, I.},
  journal    = {JOM},
  title      = {The {Materials} {Data} {Facility}: {Data} {Services} to {Advance} {Materials} {Science} {Research}},
  year       = {2016},
  issn       = {1543-1851},
  month      = aug,
  number     = {8},
  pages      = {2045--2052},
  volume     = {68},
  doi        = {10.1007/s11837-016-2001-3},
  language   = {en},
  shorttitle = {The {Materials} {Data} {Facility}},
  url        = {https://doi.org/10.1007/s11837-016-2001-3},
  urldate    = {2020-09-18},
}

@Article{Blaiszik2019,
  author   = {Blaiszik, Ben and Ward, Logan and Schwarting, Marcus and Gaff, Jonathon and Chard, Ryan and Pike, Daniel and Chard, Kyle and Foster, Ian},
  journal  = {MRS Communications},
  title    = {A data ecosystem to support machine learning in materials science},
  year     = {2019},
  issn     = {2159-6859, 2159-6867},
  month    = dec,
  number   = {4},
  pages    = {1125--1133},
  volume   = {9},
  doi      = {10.1557/mrc.2019.118},
  language = {en},
  urldate  = {2021-04-21},
}

@dataset{mdf-Cs-Rn_data,
  title={Dataset for "Accurate Atomic Correlation Energies and Virial Trends for Correlation-Consistent Effective Core Potentials III: Quantitative Separation of Fixed-Node Biases and Localization Errors in Cs–Rn"},
  author={Omar Madany and Lubos Mitas},
  publisher={Materials Data Facility},
  year={2026},
  doi={10.18126/1dd2-sd40},
  url={https://doi.org/10.18126/1dd2-sd40},
}

@article{kim2018,
doi = {10.1088/1361-648X/aab9c3},
url = {https://doi.org/10.1088/1361-648X/aab9c3},
year = {2018},
month = {apr},
publisher = {IOP Publishing},
volume = {30},
number = {19},
pages = {195901},
author = {Kim, Jeongnim and Baczewski, Andrew D and Beaudet, Todd D and Benali, Anouar and Bennett, M Chandler and Berrill, Mark A and Blunt, Nick S and Borda, Edgar Josué Landinez and Casula, Michele and Ceperley, David M and Chiesa, Simone and Clark, Bryan K and Clay, Raymond C and Delaney, Kris T and Dewing, Mark and Esler, Kenneth P and Hao, Hongxia and Heinonen, Olle and Kent, Paul R C and Krogel, Jaron T and Kylänpää, Ilkka and Li, Ying Wai and Lopez, M Graham and Luo, Ye and Malone, Fionn D and Martin, Richard M and Mathuriya, Amrita and McMinis, Jeremy and Melton, Cody A and Mitas, Lubos and Morales, Miguel A and Neuscamman, Eric and Parker, William D and Pineda Flores, Sergio D and Romero, Nichols A and Rubenstein, Brenda M and Shea, Jacqueline A R and Shin, Hyeondeok and Shulenburger, Luke and Tillack, Andreas F and Townsend, Joshua P and Tubman, Norm M and Van Der Goetz, Brett and Vincent, Jordan E and Yang, D ChangMo and Yang, Yubo and Zhang, Shuai and Zhao, Luning},
title = {QMCPACK: an open source ab initio quantum Monte Carlo package for the electronic structure of atoms, molecules and solids},
journal = {Journal of Physics: Condensed Matter},
}

@article{paul2020,
    author = {Kent, P. R. C. and Annaberdiyev, Abdulgani and Benali, Anouar and Bennett, M. Chandler and Landinez Borda, Edgar Josué and Doak, Peter and Hao, Hongxia and Jordan, Kenneth D. and Krogel, Jaron T. and Kylänpää, Ilkka and Lee, Joonho and Luo, Ye and Malone, Fionn D. and Melton, Cody A. and Mitas, Lubos and Morales, Miguel A. and Neuscamman, Eric and Reboredo, Fernando A. and Rubenstein, Brenda and Saritas, Kayahan and Upadhyay, Shiv and Wang, Guangming and Zhang, Shuai and Zhao, Luning},
    title = {QMCPACK: Advances in the development, efficiency, and application of auxiliary field and real-space variational and diffusion quantum Monte Carlo},
    journal = {The Journal of Chemical Physics},
    volume = {152},
    number = {17},
    pages = {174105},
    year = {2020},
    month = {05},
    issn = {0021-9606},
    doi = {10.1063/5.0004860},
    url = {https://doi.org/10.1063/5.0004860},
}

@article{Paul1975,
title = {Effects of relativity and of the lanthanide contraction on the atoms from hafnium to bismuth},
journal = {Chemical Physics Letters},
volume = {33},
number = {3},
pages = {408-411},
year = {1975},
issn = {0009-2614},
doi = {https://doi.org/10.1016/0009-2614(75)85741-1},
url = {https://www.sciencedirect.com/science/article/pii/0009261475857411},
author = {Paul {S. Bagus} and Yoon S. Lee and Kenneth S. Pitzer}
}

@article{jiang2010,
  title = {Electronic band structure of zirconia and hafnia polymorphs from the $GW$ perspective},
  author = {Jiang, Hong and Gomez-Abal, Ricardo I. and Rinke, Patrick and Scheffler, Matthias},
  journal = {Phys. Rev. B},
  volume = {81},
  issue = {8},
  pages = {085119},
  numpages = {9},
  year = {2010},
  month = {Feb},
  publisher = {American Physical Society},
  doi = {10.1103/PhysRevB.81.085119},
  url = {https://link.aps.org/doi/10.1103/PhysRevB.81.085119}
}

@article{foulkes2001,
  title = {Quantum Monte Carlo simulations of solids},
  author = {Foulkes, W. M. C. and Mitas, L. and Needs, R. J. and Rajagopal, G.},
  journal = {Rev. Mod. Phys.},
  volume = {73},
  issue = {1},
  pages = {33--83},
  numpages = {0},
  year = {2001},
  month = {Jan},
  publisher = {American Physical Society},
  doi = {10.1103/RevModPhys.73.33},
  url = {https://link.aps.org/doi/10.1103/RevModPhys.73.33}
}

@article{Pahl2010,
    author = {Pahl, Elke and Figgen, Detlev and Thierfelder, Christian and Peterson, Kirk A. and Calvo, Florent and Schwerdtfeger, Peter},
    title = {A highly accurate potential energy curve for the mercury dimer},
    journal = {The Journal of Chemical Physics},
    volume = {132},
    number = {11},
    pages = {114301},
    year = {2010},
    month = {03},
    issn = {0021-9606},
    doi = {10.1063/1.3354976},
    url = {https://doi.org/10.1063/1.3354976},
}

@misc{petrov2010,
      title={Correlations beyond CCSD(T) for accurate study of Hg$_2$ and Cp$_2$}, 
      author={A. N. Petrov and N. S. Mosyagin and A. V. Titov and A. V. Zaitsevskii and E. A. Rykova},
      year={2010},
  }

@article{mitas2025,
  title={Fixed-node errors in real space quantum Monte Carlo at high densities: Closed-shell atomic correlation energies},
  author={Mitas, Lubos},
  journal={Chemical Physics Letters},
  volume={860},
  pages={141770},
  year={2025},
  publisher={Elsevier}
}

@book{szabo2012,
  title={Modern quantum chemistry: introduction to advanced electronic structure theory},
  author={Szabo, Attila and Ostlund, Neil S},
  year={2012},
  publisher={Courier Corporation}
}

@article{Kuchle1997,
    author = {Küchle, W. and Dolg, M. and Stoll, H.},
    title = {Ab Initio Study of the Lanthanide and Actinide Contraction},
    journal = {The Journal of Physical Chemistry A},
    volume = {101},
    number = {38},
    pages = {7128-7133},
    year = {1997},
    month = {09},
    issn = {1089-5639},
    doi = {10.1021/jp970920c},
    url = {https://doi.org/10.1021/jp970920c},
}

@Article{ceperley1986,
author={Ceperley, D. M.},
title={The statistical error of green's function Monte Carlo},
journal={Journal of Statistical Physics},
year={1986},
month={Jun},
day={01},
volume={43},
number={5},
pages={815-826},
issn={1572-9613},
doi={10.1007/BF02628307},
url={https://doi.org/10.1007/BF02628307}
}

@article{Hammond1987,
    author = {Hammond, Brian L. and Reynolds, Peter J. and Lester, William A., Jr.},
    title = {Valence quantum Monte Carlo with ab initio effective core potentials},
    journal = {The Journal of Chemical Physics},
    volume = {87},
    number = {2},
    pages = {1130-1136},
    year = {1987},
    month = {07},
    issn = {0021-9606},
    doi = {10.1063/1.453345},
    url = {https://doi.org/10.1063/1.453345},
}
\end{document}